\documentclass[11pt]{article}    

\usepackage{graphicx}  
\usepackage{geometry}
\usepackage[russian,greek,english]{babel}
\usepackage[utf8]{inputenc}

\usepackage{bm}
\usepackage[bookmarks,breaklinks]{hyperref}  
\usepackage{titling}

\usepackage{amsmath,amsthm,amsfonts,amssymb,epsfig,color,graphics}
\newcommand{\tail}{\textit{tail}}
\newcommand{\head}{\textit{head}}
\newcommand{\bJ}{\bf J}
\newcommand{\DP}{\textsc{DP}}
\newcommand{\rahmen}[1]{\begin{center}\fbox{\parbox{11.2cm}{#1}}\end{center}}
\newcommand{\linie}{\\[-2mm]\hrule\mbox{}\\}
\newcommand{\dpc}{\textsc{PDPC}}
\newcommand{\imargin}[1]{}
\newcommand{\patchsize}{k^{2^{k}}} 
\newcommand{\fpt}{{\sf FPT}}
\newcommand{\hs}{\hspace*{0.4cm}}     
\newtheorem{obs}{Observation}
\newcommand{\tw}{\textup{\bf tw}}

\newcommand{\mso}{\textsc{msol}}

\theoremstyle{plain}

\newtheorem{theorem}{Theorem}
\newtheorem{lemma}{Lemma}

\newtheorem{proposition}{Proposition}

\begin{document}

\thanksmarkseries{alph}
\title{Planar Disjoint-Paths Completion\thanks{The second and the third author where co-financed by the European Union (European Social Fund -- ESF) and
Greek national funds through the Operational Program ``Education and Lifelong Learning'' of the
National Strategic Reference Framework (NSRF) - Research Funding Program:
``{\sl Thalis. Investing in knowledge society through the European Social Fund}''.}
\thanks{ {Emails:  Isolde Adler: {\sf iadler@informatik.uni-frankfurt.de},
Stavros Kolliopoulos: {\sf sgk@di.uoa.gr},
Dimitrios  M. Thilikos: {\sf sedthilk@thilikos.info}}}}
\author{Isolde Adler\thanks{Institut f\"{u}r Informatik, Goethe-Universit\"{a}t, Frankfurt, Germany.}
\and Stavros G. Kolliopoulos\thanks{Department of Informatics and Telecommunications, National and Kapodistrian University of Athens, Athens, Greece.}
\and Dimitrios  M. Thilikos\thanks{Department of Mathematics, National and Kapodistrian University of Athens, Athens, Greece.}\ \!\! \thanks{AlGCo project-team, CNRS, LIRMM, France.}
}

\date{}

\maketitle
\begin{abstract}
\noindent We   introduce  {\sc Planar   Disjoint  Paths   Completion},   a  completion
counterpart of the Disjoint Paths problem, and study its parameterized
complexity. The problem
can be  stated as follows: given a, not necessarily connected, plane graph $G,$ $k$ pairs of
terminals, and  a face $F$  of $G,$ find  a minimum-size
set of edges, if one exists, to be added inside $F$ so that the  embedding remains planar
and the pairs become connected  by $k$ disjoint paths in the augmented
network. Our results are twofold: first, 
we give an upper bound on the number 
of necessary additional edges when a solution exists.
This bound is a function of $k$, independent of the size of $G.$
Second, we  show that  the
problem  is fixed-parameter tractable,  in particular,  it can  be solved  in time
$f(k)\cdot n^{2}.$
\end{abstract}

{\bf Keywords}: Completion Problems, Disjoint Paths,  Planar Graphs.

\section{Introduction}

Suppose we are given a
planar road  network 
with   $n$ cities 
 and a  set of  $k$ 
 pairs of them. 
An empty area of  the network is specified and we wish to add a minimum-size
set of 
intercity roads  in that area so that 
the augmented network remains planar and the pairs are connected by $k$ internally disjoint
roads. In graph-theoretic terms, we are looking for a  minimum-size edge-completion 
of a plane graph so that an infeasible instance of the {\sc Disjoint Paths}
problem becomes feasible while maintaining   planarity.
In this paper we give an algorithm that solves  this problem 
in $f(k)\cdot n^2$ steps. Our algorithm uses a combinatorial 
lemma stating that, whenever such a solution exists, 
its size depends exclusively on $k.$

The  renowned  {\sc Disjoint Paths Problem (\DP)}  is 
defined as follows. 

\smallskip
{ 
\noindent\rahmen
{
	$\DP(G,s_{1},t_{1},\ldots,s_{k},t_{k})$
	\linie
	\textbf{Input:} An undirected  graph $G$ and $k$ pairs of 
	\emph{terminals} $s_1,t_1,\ldots,s_k,t_k\in V(G).$\\
	\textbf{Question:} are there $k$ pairwise internally vertex-disjoint
	paths $Q_1,\ldots Q_k$ in $G$ 
	such that path $Q_i$ connects $s_i$ 
	to $t_i$?\\
	(By {\em pairwise internally vertex-disjoint} we mean that two paths can only intersect at a vertex which is a terminal for both.)
}
}

\DP\  is {\sf NP}-complete even on planar graphs \cite{KramerL84thec}
but,  when  parameterized by $k,$ the problem  belongs to the parameterized complexity
class {\sf FPT}, i.e., it can be
solved in time $f(k)\cdot n^{O(1)},$ for some function $f.$  More
precisely,  
it can be solved in
$f(k)\cdot n^{3}$ time by  the celebrated algorithm of Robertson and
Seymour \cite{RobertsonS-XIII} from the Graph Minors project. For planar graphs, the same problem can be solved in $f(k)\cdot n$~\cite{ReedRSS93find}.

We introduce a completion counterpart of this problem,
{\sc Planar Disjoint Paths Completion (\dpc)}, which is of interest
on infeasible instances of \DP,  and we study its parameterized
complexity, when parameterized by $k.$ 
We are given 
an  embedding of a, possibly disconnected, planar graph $G$ in the sphere,  $k$ pairs of 
terminals $s_1,t_1,\ldots,s_k,t_k\in V(G),$   a positive integer $\ell,$
and an open
connected subset   ${\bf F}$ of  the surface of
the sphere,  such that ${\bf F}$  and $G$ do  not intersect (we stress that the boundary of ${\bf F}$ is not necessarily a cycle).
We want to determine whether  there is a set of at most $\ell$
edges   to add, the so-called {\em patch,}  so 
that \medskip

\noindent (i) the new edges lie inside ${\bf F}$ and are incident only to
vertices of $G$  on the boundary of ${\bf F},$

\noindent  (ii) the new edges do not cross with each other or with $G$, and

\noindent (iii) in
the resulting graph, which consists of $G$ plus the patch, 
\DP\ has a solution.
\medskip

\dpc \ is {\sf NP}-complete even when $\ell$ is not a part of the input and $G$ is planar by the following simple
reduction from \DP: add a triangle $T$ to $G$ and let ${\bf F}$ be the
interior of $T.$ That way, we force the set of additional edges to be 
empty and obtain $\DP$ as a special case.

Notice that our problem is polynomially equivalent to the 
minimization problem where we ask for a minimum-size patch: simply solve the problem for all possible values of $\ell.$ 
Requiring the size of the patch to be at most $\ell$ is 
the primary source of difficulty. 
In case there is no restriction on the size of the patch and we simply ask whether one exists,
the problem is in {\sf  FPT} by a reduction to   $\DP,$  which is summarized as follows. For simplicity, let ${\bf F}$ be an open disk.
Let $G'$ be the graph obtained by ``sewing" along the boundary of ${\bf F}$ an $O(n)\times O(n)$-grid. By standard arguments,
\dpc\ has a solution on $G$ if and only if \DP\ has a solution on $G'.$ A similar, but more involved,  construction applies  when
${\bf F}$ is not an open disk.

%


 
\paragraph{Parameterizing completion problems.}
Completion problems are  natural to define:  take any graph
property, represented by a collection  of graphs ${\cal P}$,
and ask whether it is possible to add edges to a graph so that the new
graph is in  ${\cal P}.$
Such problems have been studied for a long time and some of the  most prominent
are the following: 
{\sc Hamiltonian Completion} \cite[GT34]{GareyJ79comp}, {\sc Path Graph Completion}~\cite[GT36]{GareyJ79comp}
 {\sc Proper Interval Graph Completion} \cite{GolumbicKS94onth} 
 {\sc Minimum Fill-In} \cite{Yannakakis81comp}
 {\sc Interval Graph Completion} \cite[GT35]{GareyJ79comp}.  

Kaplan et al.\ in their seminal paper \cite{KaplanST99tract} initiated the study of the parameterized
 complexity of completion problems and showed that {\sc Minimum
 Fill-In}, {\sc Proper Interval Graph Completion} and {\sc Strongly
 Chordal Graph Completion} are in {\sf FPT} {\sl when parameterized by the
 number of edges to add}. 
Recently, the problem left open by \cite{KaplanST99tract}, namely {\sc Interval Graph
 Completion} was also shown to be in {\sf FPT} \cite{HeggernesPTV07inte}. 
Certainly, for all these problems the testing
 of the corresponding property is in {\sf P,}   while for problems
 such as {\sc Hamiltonian Completion}, where $\mathcal{P}$ is the class
 of Hamiltonian graphs, there is no  {\sf
 FPT} algorithm, unless {\sf P}$=${\sf NP.} 
For the same reason, one cannot expect an {\sf FPT}-algorithm
when $\mathcal{P}$ contains all YES-instances of \DP, even on planar
graphs. We consider an alternative way to parameterize
completion problems, which is appropriate for the hard case, i.e., when 
testing $\mathcal{P}$ is intractable: we  parameterize the
property itself. In this paper, we initiate this line of research, 
by considering  the parameterized property
$\mathcal{P}_k$ that contains all YES-instances of \DP\  on planar
graphs with $k$ pairs of
terminals. 
%
%
%
%
%
%
%
%
%
%
%


\paragraph{Basic concepts.} As open sets are not discrete structures, we introduce some formalism
that will allow us to move seamlessly from   topological  to combinatorial arguments.
The definitions may look involved at first reading, but this is warranted  if one  considers, as we do,  the problem in its full
generality where the input  graph is not necessarily connected.





Let $G$ be a graph embedded in the sphere ${\rm \Sigma}_{0}.$ 
Embeddings are always without crossings and we view the 
graph $G$ as
a  subset of ${\rm \Sigma}_{0}.$
Given a set ${\bf X}\subseteq {\rm \Sigma}_{0}$,
let ${\bf clos}({\bf X})$, ${\bf int}({\bf X})$, and $\partial{\bf X}$ denote 
the closure, the interior,  and the boundary of $X,$  respectively.
We define $V({\bf X})=V(G)\cap {\partial}{\bf X}.$ 
A {\em noose} is a Jordan curve of ${\rm \Sigma}_{0}$ that meets $G$ only on  vertices. Let ${\cal D}$ be a finite collection
of mutually non-intersecting open disks of ${\rm \Sigma}_{0}$ whose boundaries are nooses
and such that each point that belongs to at least two such nooses is a vertex of $G.$ 
We define ${\bf I}_{{\cal D}}=\bigcup_{D\in {\cal D}}D$ 
and define ${\rm \Gamma}_{{\cal D}}$ as the 
${\rm \Sigma}_{0}$-embedded graph 
whose vertex set is $V({\bf I}_{\cal D})$
and whose edge set consists of the connected components 
of the set $\partial {\bf I}_{\cal D}\setminus V({\bf I}_{\cal D}).$
Notice that, in the definition of ${\rm \Gamma}_{\cal D}$, we permit multiple edges, loops, or vertex-less edges.

Let ${\bf J}$ be an open subset of ${\rm \Sigma}_{0}.$ 
${\bf J}$ is a {\em cactus set} of $G$ if there is a collection ${\cal D}$ as above such that  ${\bf J}={\bf I}_{{\cal D}}$,
 all biconnected components of the graph ${\rm \Gamma}_{\cal D}$ are cycles, $V(G)\subseteq  {\bf clos}({\bf J})$, and
each edge of  $E(G)$ is a subset of ${\bf J}.$ 
 Given such a ${\bf J}$, we define
 ${\rm \Gamma}_{\bf J}={\rm \Gamma}_{\cal D}.$~%
 Two cactus sets  ${\bf J}$ and ${\bf J'}$ of
 $G$ are {\em isomorphic} if ${\rm \Gamma}_{\bf J}$ and ${\rm \Gamma}_{\bf J'}$ are
 topologically isomorphic. Throughout this paper, we use the standard 
 notion of topological isomorphism between planar embeddings, 
see Section~\ref{sec:prel}.
 %
 %
%
 
\begin{figure}[h]
\label{prelexamples}
\begin{center}
\scalebox{.06}{\includegraphics{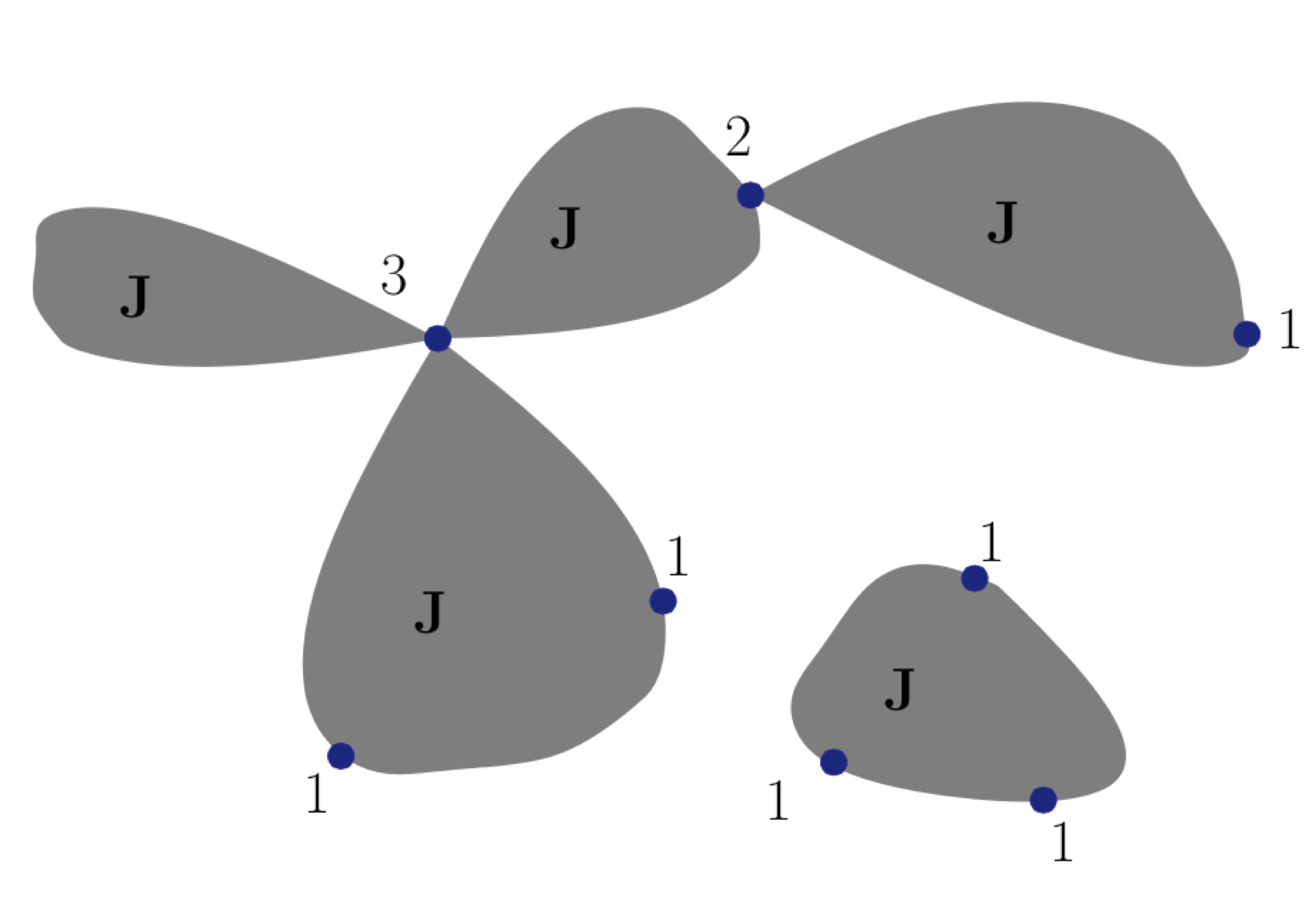}}
\end{center}
\caption{\small A cactus set ${\bf J}$ and the vertices of $V({\bf J}).$ Next to each vertex $v$ we give its multiplicity $\mu(v).$}
\end{figure}

Given a cactus set ${\bf{J}}$, we define for each vertex $v\in V({\bf J})$
its {\em multiplicity} $\mu(v)$ to be equal to the number of connected components 
of the set ${\rm \Gamma}_{\bf J}\setminus \{v\}$, minus the number of connected components of ${\rm \Gamma}_{\bf J},$ plus one. We also define $\mu({\bf J})=\sum_{v\in V({\bf J})}\mu(v).$
Observe that, given a cactus set ${\bf J}$ of $G,$ the edges of $G$ lie 
entirely within the interior of ${\bf J}.$ See Figure~\ref{prelexamples}.
\imargin{why? And: Do we need this?}
The boundary of  ${\bf J}$ corresponds to a collection of
simple closed curves such that 
\begin{itemize}
\item[(i)] no two of them intersect at more
than one point and 
\item[(ii)] they intersect with $G$ only at (some of) the
vertices in $V(G).$ 
\end{itemize}
 Cactus sets are useful throughout our paper as
``capsule'' structures that surround $G$ and thus
they abstract the interface of a graph embedding with 
the  rest of the sphere surface. 

\begin{figure}[t]
\begin{center}
\scalebox{.222}{\includegraphics{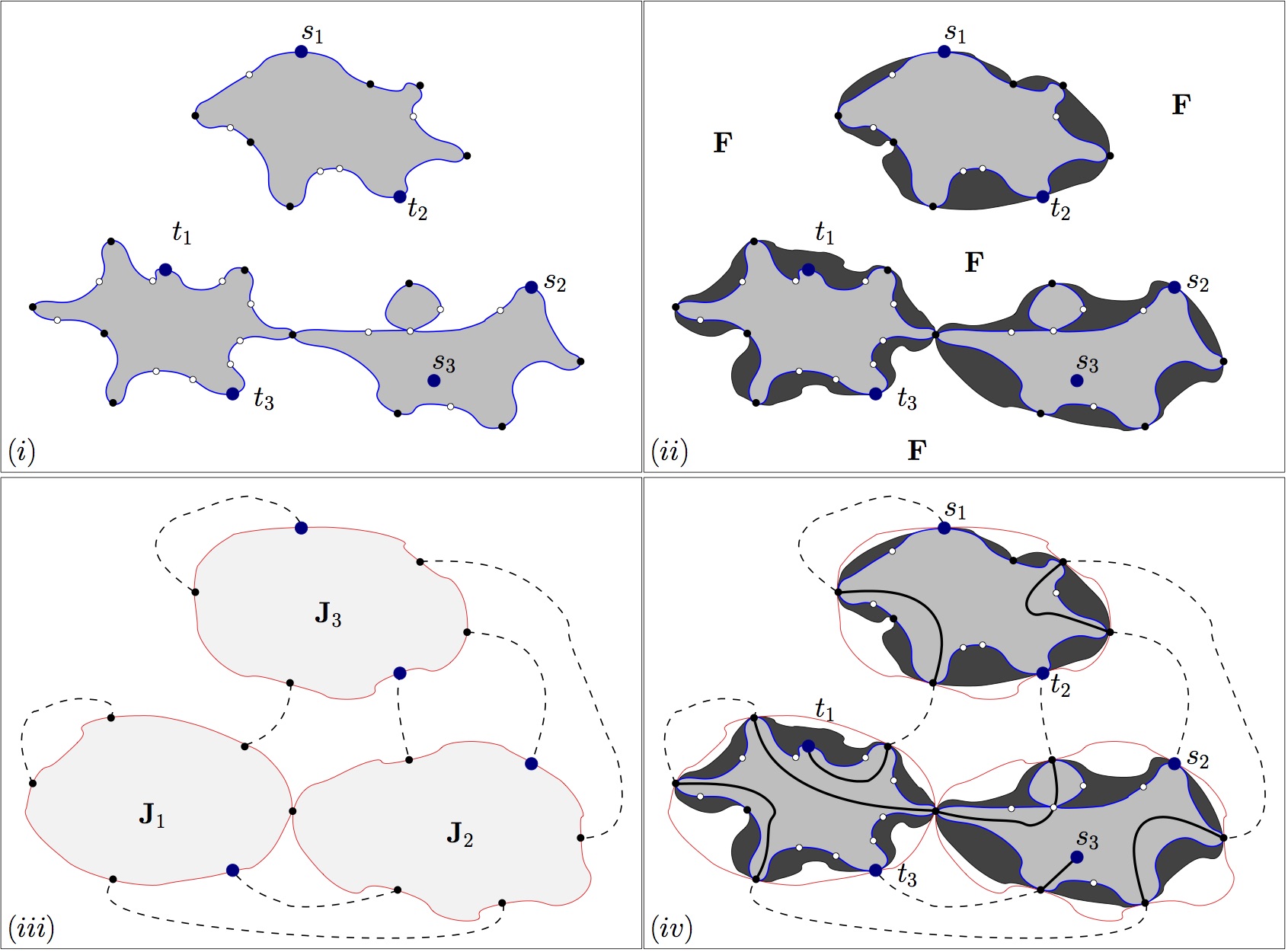}}
\end{center}
\caption{\small An example  input of the \dpc\ problem and a solution to it when $\ell=8$: $(i)$ The graph embedding in the input and the terminals $s_{1},t_{1},s_{2},t_{2},s_{3},t_{3}.$ 
The closure of the grey area contains the graph $G$ and the big vertices are the terminals. 
The white area is a face of $G.$
$(ii)$ The input of the problem, consisting of $G$, the terminals and
the outer-cactus set ${\bf F}.$ The solid black vertices  are the vertices
of $G$ that are also vertices of $V({\bf F}).$ $(iii)$ The solution of the problem
consists of the ${\bf F}$-patch $(P,{\bf J})$ where the edges of $P$
are the dashed lines and ${\bf J}={\bf J}_{1}\cup{\bf J}_{2}\cup
{\bf J}_{3}.$  $(iv)$ The input and the solution together where
the validity of the patch is certified by 3
disjoint paths. 
}
\label{wefig:example}
\end{figure}

 We  say that an open set ${\bf F}$ of ${\rm \Sigma}_{0}$
 is an {\em  outer-cactus set} of $G$ if ${\rm \Sigma}_{0}\setminus{\bf clos}({\bf F})$ is a cactus set of $G.$
 See Fig.~\ref{wefig:example}.($ii$). For example, if $G$ is planar, any face
 $F$ of $G$ can be used to define an outer-cactus set, whose boundary
 meets $G$ only at the vertices incident to $F.$ 
 Our definition of an
 outer-cactus set is more general:   it can  be a subset of a face $F,$ meeting the
 boundary of $F$ only at some of its vertices. 

 Let $G$ be an input graph to \DP, see Figure~\ref{wefig:example}.($i$). Given an outer-cactus set ${\bf F}$ of $G$, 
an ${\bf F}$\emph{-patch} of $G$ is a pair $(P,{\bf J})$
 where 
 (i) ${\bf J}$ is a cactus set of $G$, where ${\rm \Sigma}_{0}\setminus
 {\bf clos}({\bf J})\subseteq {\bf F}$
and (ii) $P$ is a graph embedded in ${\rm \Sigma}_{0}$ {\sl without crossings}
such that 
$E(P)\subseteq {\rm \Sigma}_{0}\setminus {\bf clos}({\bf J})$, $V(P)=V({\bf
J})$ (see Figures~\ref{wefig:example}.($iii$) and~\ref{wefig:example}.($iv$)).
Observe that the  edges of $P$ do not cross any edge in $E(G).$ 
In the definition of the ${\bf F}$-patch, 
the graph $P$ contains the new edges we add. 
The vertices in $V({\bf F})$
define the vertices of $G$  which we are allowed to include in
$P.$ 
$V({\bf J})$ is meant to contain those vertices of $V({\bf F})$ 
that become vertices (possibly isolated) of $P.$ 
%
%
%
%
%
In terms of data structures, we assume that a 
 cactus set ${\bf J}$ is represented by  the (embedded) graph ${\rm \Gamma}_{{\bf
J}}.$  Similarly an outer-cactus set ${\bf F}$ is represented by
the (embedded) graph ${\rm \Gamma}_{\Sigma_0 \setminus {\bf clos({\bf F})}}.$

%
%

%
%
%
%
%
%
We restate now the definition of the {\sc Planar Disjoint Paths
Completion} problem as follows: 

\noindent\rahmen
{\small
	$\dpc(G,s_{1},t_{1},\ldots,s_{k},t_{k},\ell,{\bf F})$
	\linie
	\textbf{Input:} A graph $G$ embedded  in ${\rm \Sigma}_{0}$ without 
        crossings,		
	terminals $s_1,t_1,\ldots,s_k,t_k\in V(G),$ a positive integer $\ell$, and an outer-cactus
	set ${\bf F}$ of $G.$ \\
	\textbf{Parameter:} $k$ \\	
	\textbf{Question:} Is there an {\bf F}-patch $( P ,{\bf J})$ of $G$, such that $|P|\leq \ell$ and
	$\DP(G\cup  P ,s_1,t_1,\ldots,s_k,t_k)$ has a solution?
	Compute such an {\bf F}-patch if it exists.
	
	} 
	\imargin{how are we given the embedding? }\imargin{I see that 
	an outer-cactus set {\bf F}
	might not contain all possible vertices on the boundary. Correct?}

\noindent If such an ${\bf F}$-patch exists, we call it a \emph{solution} for
$\dpc.$ In the corresponding optimization problem, denoted by {\sc min}-\dpc, one asks for the minimum $\ell$ for which  $\dpc$ has a solution, if one exists.
See Fig.~\ref{wefig:example} for an example input of \dpc\ and 
a solution to it. 

%
%

%
%
%
%
%

\paragraph{Our results.}
Notice that in the definition of \dpc\   the size of the patch does not depend on the parameter $k.$ Thus, it is not even obvious that $\dpc$ belongs to the parameterized complexity  class {\sf XP,}
i.e., it has an algorithm of time $n^{f(k)}$ for some function $f.$  
Our first contribution, Theorem~\ref{theo:patchsize}, is a combinatorial one: we prove that  if a patch exists, then its size is bounded by $\patchsize.$  Therefore, we can always assume that $\ell$ is bounded 
by a function of $k.$
This bound is the departure point for the proof of the main 
algorithmic result of this paper:

\begin{theorem}\label{theo:main}
	$\dpc\in\fpt.$ In particular, $\dpc$ can be solved in $f(k)\cdot n^{2}$ steps, where $f$ is a function
	that depends only on $k.$ Therefore, {\sc min}-\dpc\ can be solved in $g(k)\cdot n^{2}$ steps.
\end{theorem}

\noindent 
We present now the proof strategy and the ideas
underlying our results.

\subsection{Proof strategy} 

\paragraph{Combinatorial Theorem.}   In Theorem~\ref{theo:patchsize}, we prove that every patch whose size is larger than $\patchsize,$
can be replaced by another one of strictly smaller size. In
particular, we identify 
a region ${\bf B}$ of ${\bf F}$ that is traversed  by a large number of segments of different paths of the \DP\ solution. Within that region, we apply a global 
topological transformation that replaces the old patch by a new, strictly smaller one, while preserving its embeddability in ${\bf F}.$ The planarity of the new patch 
is based on the fact that the new segments are  reflections in  ${\bf
  B}$ of a set of segments of 
the feasible \DP\ solution that previously lied outside ${\bf B}.$ This combinatorial result allows us 
to reduce the search space of the problem to one whose size is bounded by $\min\{\ell, \patchsize\}.$ Therefore, the construction of the corresponding collection of 
 ``candidate  solutions'' can be done in advance, for each given $k$, without requiring {\sl any a priori knowledge} of the 
 input graph $G.$ 
 
 We note that the proof of our combinatorial  theorem could be of independent interest. 
 In fact, it is one of the ingredients of the proof of the main  result of~\cite{AdlerKKLST12tigh}.

\paragraph{The algorithm for $\dpc.$}  As the number of patches 
is bounded by a function of $k,$ we need to determine
whether  there is a correct way to glue one of them 
on vertices of the boundary of the open set ${\bf F}$ 
so that the resulting graph is a YES-instance of the \DP\ problem. For
each candidate patch $\tilde{P},$ together with its corresponding
candidate  cactus set  $\tilde{\bf J}$, we define the {\sl set of compatible}  
 graphs embedded in $\tilde{\bf J}.$ Each  compatible
graph $\tilde{H}$ consists of unit-length paths and
has the property that $\tilde{P}\cup \tilde{H}$ contains $k$ disjoint paths. Intuitively, each $\tilde{H}$ 
is a certificate of the part of the \DP\ solution 
that lies within $G$ when the patch in ${\bf F}$ 
is isomorphic to $\tilde{P}.$ It therefore remains 
to check for each $\tilde{H}$ whether 
it can be realized by a collection of actual paths within $G.$ For this, we set up a collection ${\cal H}$ 
of all such certificates.
Checking for a suitable realization of a member of ${\cal H}$ in $G$
is still a topological problem that depends on the embedding of $G$: graphs that are 
isomorphic, but not {\sl topologically isomorphic,} 
may certify different completions. 
For this reason, our next step 
is to {\sl enhance} the structure of the members of ${\cal H}$   
so that their realization in $G$ reduces to a purely combinatorial
check. (Cf. Section~\ref{sec:enhance}   for the
definition of the enhancement operation). 
We show in Lemma~\ref{moster} that for the enhanced certificates,
this check can be  implemented by  rooted   topological minor testing.
For this check, we can apply the recent algorithm of~\cite{GroheKMW10find}
that runs in $h_{1}(k)\cdot n^{3}$ steps
and obtain an algorithm of overall complexity $h_{2}(k)\cdot n^{3}.$

We note that the use of the complicated machinery of the algorithm in~\cite{GroheKMW10find}
can be bypassed towards obtaining a simpler and faster  $f(k)\cdot n^{2}$ algorithm.
This is possible because the generated instances of the rooted topological minor problem 
satisfy certain structural properties.  This allows the direct application of the {\sl Irrelevant Vertex Technique}
introduced in~\cite{RobertsonS-XIII} for solving, among others, the {\sc Disjoint Paths} Problem. 
The details of this improvement are in Section~\ref{sec:irrel}.
\section{Preliminaries}
\label{sec:prel}
We consider finite graphs. For a graph $G$ we denote the vertex set by $V(G)$ and the edge set by $E(G).$ If $G$ is embedded  in the sphere ${\rm \Sigma}_{0}$,  
the edges of $G$ and the graph $G$ refer also to the corresponding sets of points in $
{\rm \Sigma}_{0}.$ Clearly the edges of $G$ correspond to open sets and $G$ itself is a closed set. We denote by $F(G)$ the set of all the faces of $G$, i.e., all connected components of ${\rm \Sigma}_{0}\setminus G.$
Given a set $S\subseteq V(G),$ we 
say that the pair $(G,S)$ is a graph {\em rooted} at $S.$ We also denote as ${\cal P}(G)$ the set of all paths in $G$ with at least one edge.
Given a path $P\in{\cal P}(G)$, we denote by $I(P)$ the set of internal vertices of $P.$
Given a  vertex $v\in V(G)$ and a positive integer $r$, we denote  by $N^{r}_{G}(v)$ the set of all vertices in $G$ that are within distance at most $r$ from $v.$
Given a vertex $v$ of a graph $G$ with exactly two neighbors $x$ and $y$,
the result of the {\em dissolution} of $v$ in $G$ is the graph obtained 
if we remove $v$ from $G$ and add, if it does not already exist, the edge $\{x,y\}$.
\medskip

\noindent {\bf Rooted topological minors.}
Let $H$ and $G$ be graphs, $S_{H}$ be a subset 
 of vertices in $V(H)$, $S_{G}$ be a subset 
 of vertices in $V(G)$, and  $\rho$
be a bijection from $S_{H}$ to $S_{G}.$
We say 
that $(H,S_{H})$ is a {\em $\rho$-rooted topological minor} of $(G,S_{G})$, if there exist
injections $\psi_{0} \colon V(H)\rightarrow V(G)$ and $\psi_{1} \colon
E(H)\rightarrow {\cal P}(G)$ such that \\
\\   %
\hs  ${\bf 1.}$  $\rho\subseteq \psi_{0}$,   \\
\hs ${\bf 2.}$ for  every $e=\{x,y\}\in E(H)$, $\psi_{1}(e)$ is a $(\psi_{0}(x),\psi_{0}(y))$-path in ${\cal P}(G)$, and  \\
\hs  ${\bf 3.}$ all  $e_{1},e_{2} \in E(H)$ with $e_1 \neq e_2$
satisfy $I(\psi_{1}(e_{1}))\cap V(\psi_{1}(e_{2}))=\emptyset.$ \\

In words, when $H$ is a topological minor of $G,$ 
$G$ contains  a subgraph which is isomorphic to 
a subdivision of $H.$ In addition, when  $(H,S_{H})$ is a {\em $\rho$-rooted topological minor} of $(G,S_{G})$ then this isomorphism respects the bijection $\rho$ between the vertex sets $S_{H}$ and $S_{G}.$
\medskip

\noindent {\bf Contractions.}
Let $G$ and $H$ be graphs and let $\sigma \colon V(G)\rightarrow V(H)$ be
a surjective mapping  such that
\\ 

\noindent \hs  ${\bf 1.}$  for every vertex $ v\in V(H)$, the graph $G[\sigma^{-1}(v)]$ is connected;  \\
\hs  ${\bf 2.}$ for every edge $ \{v,u\}\in E(H)$, the graph
$ G[\sigma^{-1}(v)\cup \sigma^{-1}(u)]$ is
connected;  \\
\hs ${\bf 3.}$ for every $ \{v,u\}\in E(G)$,  either $\sigma(v)=\sigma(u)$,
or $\{\sigma(v),\sigma(u)\}\in E(H).$\\

\noindent
We say that {\em $H$ is a $\sigma$-contraction of $G$} or simply that $H$ is a {\em contraction} of $G$ if such a $\sigma$ exists.

\begin{obs}
\label{obs:distsl}
Let $H$ and $G$ be graphs such that $H$ is a $\sigma$-contraction of $G.$ If $x,y\in V(G)$, then the distance in $G$ between $x$ and $y$ is at least the distance in $H$ 
between $\sigma(x)$ and $\sigma(y).$
\end{obs}

%
%
%





We also need the following topological lemma.

\begin{lemma}
\label{topolemma}
Let $G$ be a ${\rm \Sigma}_{0}$-embedded  graph and let ${\bf J}$ be a cactus set of it. Let 
also $M$ be a ${\rm \Sigma}_{0}$-embedded graph such that $M\cap {\bf J}=\emptyset$ and $V(M)\subseteq V({\bf J}).$
Then there is a closed curve $K$ in $\Sigma_{0}\setminus {\bf clos}({\bf J})$ meeting each edge of $M$ twice.
\end{lemma}

\paragraph{Proof.}
We consider the dual graph of ${\rm \Gamma}_{\bf J}\cup M$ and we remove from it all vertices lying inside ${\bf J}.$ We denote by $Q$ the resulting graph and notice that 
$Q$ is connected because the set ${\rm \Sigma}_{0}\setminus {\bf clos}({\bf J})$ is connected. Next, 
construct the graph $Q'$ by subdividing once 
every edge of $Q$ such that the subdivision vertex
 is the intersection point of the edge and its dual. 
Let $T$ be a spanning tree of $Q'$ 
and let $K$ be a closed curve such that $T$ is inside one of the  connected components of the set ${\rm \Sigma}_{0}\setminus K$
and ${\bf J}$ in the other.  If we further require $K$ to intersect
$M$ a minimum number of times, we obtain the claimed curve.\qed
\medskip

\noindent {\bf Topological isomorphism.} 
Given a graph $G$ embedded in ${\rm \Sigma}_{0}$, let ${\bf f}$ be a face in $F(G)$ whose boundary has $\xi$ connected components $A_{1},\ldots,A_{\xi}.$ 
We define the set $\pi({\bf f})=\{\pi_{1},\ldots,\pi_{\xi}\}$ 
such that each $\pi_{i}$ is the cyclic ordering of $V(A_{i})$, possible with repetitions, defined by the way vertices are met while walking along $A_{i}$
in a way that the face ${\bf f}$ is always on our left side. Clearly, repeated vertices 
in this walk are cut-vertices of $G.$

Let $G$ and $H$ be graphs embedded
in ${\rm \Sigma}_{0}.$ We say that $G$ and $H$ are {\em topologically
isomorphic} if there exist bijections $\phi \colon  V(G)\rightarrow V(H)$ 
and $\theta \colon F(G)\rightarrow F(H)$ such that \medskip
\\ \hs  ${\bf 1.}$  $\phi$ is an isomorphism from $G$ to $H$, i.e. 
for every pair $\{x,y\}$ of distinct vertices  in $V(G)$,
$\{x,y\}\in E(G)$ iff $\{\phi(x),\phi(y)\}\in E(H).$ \\
\hs  ${\bf 2.}$  For every face ${\bf f}\in F(G)$, $\phi(\pi({\bf f}))=\pi(\theta({\bf f})).$
\medskip

\noindent
In the definition above, by $\phi(\pi({\bf f}))$ we mean $\{\phi(\pi_{1}),\ldots,\phi(\pi_{\xi})\}$, where, if $\pi_{i}=(x_{1},\ldots,$\ $x_{\zeta_{i}},x_{1})$, then by $\phi(\pi_{i})$, we mean $(\phi(x_{1}),\ldots,\phi(x_{\zeta_{i}}),\phi(x_{1})).$
Notice that it is possible for two isomorphic planar graphs to have embeddings 
that are not topologically isomorphic (see~\cite[page 93]{Diestel00grap} for such an example and further discussion on this topic).\medskip
 
\noindent {\bf Treewidth.}
A \emph{tree decomposition} of a graph $G$ is a pair $(\mathcal{X},T)$ where $T$
is a tree with nodes $\{1,\ldots,m\}$ and ${\cal X}=\{X_{i} \mid i\in V(T)\}$ is a collection of subsets
of $V(G)$ (called \emph{bags}) such that: \\
\\ \hs  ${\bf 1.}$ $\bigcup_{i \in V(T)} X_{i} = V(G)$, \\
\hs  ${\bf 2.}$ for each edge $\{x,y\} \in E(G)$, $\{x,y\}\subseteq X_i$ for some
\hs  $i\in V(T)$, and  \\
\hs  ${\bf 3.}$ for each $x\in V(G)$ the set $\{ i \mid x \in X_{i} \}$
induces a connected subtree of $T.$\\

\noindent 
The \emph{width} of a tree decomposition $(\{ X_{i} \mid i \in V(T) \},
T)$ is $\max_{i \in V(T)}\,\{|X_{i}| - 1\}.$ The \emph{treewidth} of a
graph $G$ denoted $\tw(G)$ is the minimum width over all tree decompositions of $G.$

\section{Bounding the size of the completion}\label{sec:bounding-fill-in}

In this section we show the following.
\begin{theorem}\label{theo:patchsize}
	If there is a solution for $\dpc(G,s_1,t_1,\ldots,s_k,t_k,\ell,{\bf F})$,
	then there is a solution 
	$(P,{\bf J} )$ with $\left|E( P )\right|\leq \patchsize  .$ 
\end{theorem}

For the proof, we need the following combinatorial lemma.

\begin{lemma}\label{lem:colorsequence}
	Let $\Sigma$ be an alphabet of size 
	$\left|\Sigma\right|=k.$ 
	Let $w\in\Sigma^*$ be a string over $\Sigma.$ 
	If $\left| w \right|>2^k$, then 
	$w$ contains an infix $y$ with $\left| y \right|\geq2$,  
	such that every letter occurring in $y$, 
	occurs an even number of times in $y.$
\end{lemma}

\paragraph{Proof.}
Let $\Sigma=\{a_1,\ldots,a_k\}$, and
let $w=w_1\cdots w_n$ with $n>2^k.$ Define vectors
$z_i\in \{0,1\}^k$ for $i\in\{1,\ldots, n\}$, and we let the $j$th entry
of vector $z_i$ be $0$ if and only if letter $a_j$ occurs an even number
of times in the prefix $w_1\cdots w_i$ of $w$ and $1$ otherwise.
Since $n>2^k$, there exist $i,i'\in\{1,\ldots, n\}$ with $i\neq i'$,
such that $z_i=z_{i'}.$
Then $y=w_{i+1}\cdots w_{i'}$ proves the lemma.\qed\medskip

We also need the following easy topological lemma.

\begin{lemma}
\label{standard}
Let $G$ be a connected outerplanar graph  that may have loops but no multiple edges  (or loops)  and let $\Gamma$ be an embedding of $G$ in $\Sigma_{0}$ such that 
all its vertices belong to the boundary of a face $f$.
Let also $E_{\rm in}$ be the set of  all the edges or loops of $G$ that do not belong to the boundary of $f$. 
Then there is a unique (up to topological isomorphism) embedding $\Gamma'$ of $G$ such that $\Gamma\cap \Gamma'$ is the boundary of $f$ and that all edges of $E_{\rm in}$ are embedded inside $f$.
\end{lemma}

\proof
Let $\tilde{\Gamma}$ be the boundary of $f$.
Let ${\cal F}$ be the set of all connected components of $\Sigma_{0}\setminus \tilde{\Gamma}$, except from $f$.

For each $e\in E_{\rm in}$ we denote by $S_{e}$ the  set of endpoints of $e$.
For each edge $e\in E_{\rm in}$ we define a set $\bar{e}$ that will be the image of $e$ in the new embedding $\Gamma'$.
We distinguish the following two cases.\smallskip

\noindent{\em Case 1}. $|S_{e}|=2$. Notice that $S_{e}$ is a separator 
of $G$. Notice also that $G\setminus S_{e}$ contains two connected 
components. Also there exists a (unique)  cycle $C_{e}$ containing the edge $e$
and no other points of $\Gamma$
such that each of these connected  components belong each to a distinct  disk among the two disks that are bounded by $C_{e}$.
We define $\bar{e}=C_{e}\setminus {\bf clos}(e)$.\smallskip

\noindent{\em Case 2}.   $|S_{e}|=1$. Notice that there exists a unique
cycle $C_{e}$ defining two disks $D$ and $D'$ such that (i)
$C_{e}\cap \Gamma=S_{e}$ 
and (ii)  $\Gamma_{e}\subseteq D$ where $\Gamma_{e}$ is the  (unique)
member of ${\cal F}$ that contains the loop $e$. 
We define $\bar{e}=C_{e}\setminus S_{e}$.\smallskip

\noindent Notice that it is possible to define all $\bar{e}$'s, one by one, such 
that they do not intersect.
Given the above definitions, it follows that 
$\tilde{\Gamma}\cup\bigcup_{e\in E_{\rm in}}{\bar{e}}$
is the required embedding. 
\qed

%
%

\medskip\noindent\emph{Proof  of Theorem~\ref{theo:patchsize}}.
Let $(P,\bJ) $ be a solution for $\dpc(G,s_1,t_1,\ldots,s_k,t_k,\ell,{\bf F})$
with $\left|E( P )\right|$ minimal.
Consider the embedding of  $G\cup P$ in the sphere $\Sigma_0$, and 
let $Q_1,\ldots,Q_k$ be the paths of a \DP\ solution in $G\cup P .$
By the minimality of $\left|E( P )\right|$ we can assume that  the edges of 
$P$ are exactly the edges of $\bigcup_{i\in\{1,\ldots,k\}}Q_{i}$ that are not in $G.$ For the same reason, two edges in $P$ have a common endpoint $x$ that is not a terminal {\sl only if}
$x$ is a cut-vertex of ${\rm \Gamma}_{\bf J}.$

Let $P^*$ denote the graph obtained from the dual of $P\cup {\rm \Gamma}_{\bf J}$,
after removing the vertices corresponding to the faces of ${\rm \Gamma}_{\bf J}$ that are disjoint from ${\bf F}.$
We first show that the maximum degree of $P^*$ is bounded by $k.$
Assume, to the contrary, that $P^{*}$ has a vertex incident to two edges $e^*_{1}$ and $e^{*}_{2}$ 
such that $e_{1}$ and $e_{2}$ belong
to the same path, say $Q_{x}$, in $\{Q_{1},\ldots,Q_{k}\}.$ 
Then  it is possible to choose an endpoint $p_{1}$ of $e_{1}$ 
and an endpoint $p_{2}$ of $e_{2}$ such that, $(P',\bJ ')$ is also a solution  such that $|E(P')|<|E(P)|.$ Indeed, $p_1$ and $p_2$ are the vertices incident to $e_1$ and $e_2$ belonging to the two connected components of $Q_x \setminus  \{e_1, e_2\}$ that contain respectively $s_x$ and $t_x$. Therefore we established that for every $i \in  [k]$, no vertex of $P^*$ is incident to two distinct edges of $Q_i$, thus 
all vertices of  $P^*$ have degree at most $k$.

Here  $P'=P\setminus\{e_{1},e_{2}\}\cup\{\{p_{1},p_{2}\}\}$ and ${\bf J}'$ is defined  such that ${\rm \Gamma}_{{\bf J}'}$ is obtained from ${\rm \Gamma}_{\bf J}$ after dissolving the vertices  that became isolated during the construction of $P'$.

Our next aim is to prove that 
 the diameter of $P^*$ is bounded by $2^k.$ Then
$\left|E(P^*)\right|=\left|E(P)\right|\leq k^{2^{k}}$ and we are done.
\imargin{possibly not optimal: haven't used that 
$P$ is a collection of cycles and paths}
Note that every edge in $E( P ^*)$
corresponds to an edge in exactly one path of $Q_1,\ldots,Q_k.$
Hence, given a path $R=(r_0,\ldots ,r_{\zeta})$ in $P^*$, it corresponds to a 
string $w\in \{Q_1,\ldots, Q_k\}^*$ in a natural way. 
It is enough to prove the following claim.\medskip

\noindent{\bf Claim.}
{\em The string $w$ contains no infix $y$ with $\left| y \right|\geq2$,  
such that every letter occurring in $y$ 
occurs an even number of times in $y.$}
\smallskip

\paragraph{Proof of the Claim.}
Towards a contradiction, suppose that $w$ contains such an infix $y.$
We may assume that $w=y.$
Let $E_R\subseteq E( P )$ be the set of edges corresponding to the edges of the
path $R\subseteq P^*.$  Then $\left|E_R\right|\geq 2$ because $w$ (and hence $R$) has length
at least $2.$ Let ${\bf B}\subseteq \Sigma_0$ be the open set defined 
by the union of all edges in $E_{R}$ and all faces of the graph $P\cup {\rm \Gamma}_{\bf J}$ that are incident to them. Clearly, ${\bf B}$ is an open  connected subset of ${\bf F}$ with the following properties: 
\begin{itemize}
\item[(a)] ${\bf B}$ contains  all  edges in $E_R$ and no other edges of $P$, 
\item[(b)] the ends of every edge in $E_R$ lie on the boundary $\partial {\bf B}$, and
\item [(c)]
every edge in $E(P)\setminus E_R$ has empty intersection with ${\bf B}.$
\end{itemize}
We consider an `up-and-down' partition
$(U=\{u_{1},\ldots,u_{r}\},D=\{d_{1},\ldots,d_{r}\})$ 
\imargin{why not take $U$ and $D$ to be tuples?}
of the endpoints of the
edges in $E_{R}$  as follows: traverse the path $R$ in $P^{*}$ in some
arbitrary direction and when the $i$th edge $e_{i}\in E_{R}$ is met, the
endpoint $u_{i}$ of $e_{i}$ \imargin{$e_i$?} 
on the left of this direction is added to $U$ and the
right endpoint $d_{i}$ is added to $D.$ Notice that $U$ and $D$ may be
multisets because it is not necessary that  all vertices in $P$ have degree $1$. 
For each $i\in\{1,\ldots,r\}$ we say that $u_{i}$ is the {\em
counterpart} of $d_{i}$ and vice versa.\medskip

\noindent Because the paths $Q_1,\ldots,Q_k$ are vertex-disjoint, the following holds. 
 
 \noindent{\sl Observation:}
If $x\in V(P)$ has degree larger than one, then either $x$ is a terminal and 
has degree at most $k$ or $x$ is a cutpoint of ${\rm \Gamma}_{\bf J}$ and  has 
degree exactly $2$.

\medskip

\begin{figure}[t]

\begin{center}
\scalebox{.24}{\includegraphics{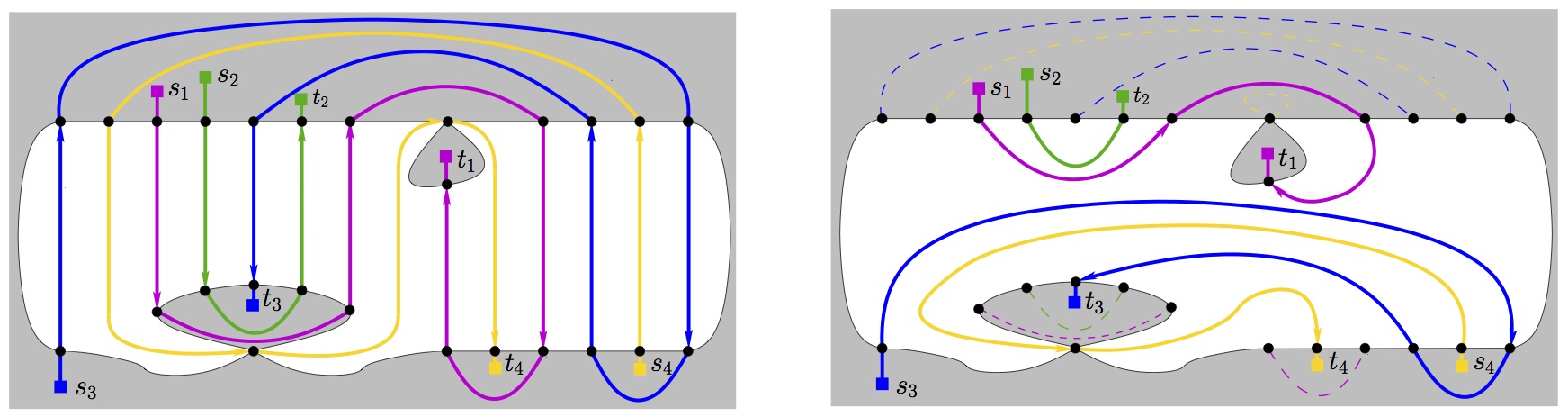}}
\end{center}
\caption{Example of the transformation in the proof of the Claim in the proof of Theorem~\ref{theo:patchsize}; $P$ is on the left and $P'$ is shown on the right. The dashed lines represent the edges of $C.$ 
}
\label{fig:reflection}
\end{figure}

By assumption, every path $Q_i$ crosses $R$ an even, say $2n_{i}$,  number of times.  
Now for every
path $Q_i$ satisfying $E(Q_i)\cap E_R\neq\emptyset$, we number the edges in
$E(Q_i)\cap E_R$ by $e^i_1,\ldots,e^i_{2n_i}$ in the order of their appearance when
traversing $Q_i$ from $s_i$ to $t_i$ and we orient them from $s_i$ to $t_i.$
We introduce shortcuts for $Q_i$ as follows: for every odd number
$j\in\{1,\ldots,2n_i\}$, we replace the subpath of $Q_i$ from $\tail(e^i_j)$ to
$\head(e^i_{j+1})$ by a new edge $f^i_j$ in $D$ (see Figure~\ref{fig:reflection}).\imargin{in $B$}

After having done this for all odd
numbers $j\in\{1,\ldots,2n_i\}$, we obtain a new path $Q_i'$ from $s_i$ to
$t_i$ that uses strictly less edges in ${\bf B}$ than $Q_i.$ Having replaced all
paths $Q_i$ with $E(Q_i)\cap E_R\neq\emptyset$ in this way by a new path
$Q_i'$, we obtain from $P$ a new graph $P '$ by replacing every pair of edges
$e^i_j,e^i_{j+1}\in E( P )$ by $f^i_j$ for all $i\in\{1,\ldots,k\}$ with
$E(Q_i)\cap E_R\neq\emptyset$, and for all $j\in \{1,\ldots,2n_i\}$, $j$ odd. 
We denote by $E_{R}'$ the set of all replacement edges $f^{i}_{j}.$
We also remove every vertex that becomes isolated in $P'$ during this operation. 

Then it is easy to verify that:
\begin{itemize} \item None of the edges of $E_{R}$ survives in $E(P').$
\item $\left|E( P ')\right|<\left|E( P )\right|.$
	\item $\DP(G\cup  P ',s_1,t_1,\ldots,s_k,t_k)$ has a solution.
\end{itemize} 
If we show that, for some suitable cactus set ${\bf J}'$ of $G$,  $(P ',{\bf J}')$ is an {\bf F}-patch, then we are done, 
because $|E(P')|<|E(P)|.$ 
%
In what follows, we prove that $ P '$ can also be embedded without crossings in ${\bf clos}({\bf F})$ such that $E(P')\subseteq {\rm \Sigma}_{0}\setminus \partial{\bf F}.$ For this it suffices to prove that 
the edges in $E_{R}'$ can be embedded in ${\bf B}$ without crossings.



%

For every path $Q_i$ with $E(Q_i)\cap E_R\neq\emptyset$ let $F^i_j$ denote the
subpath of $Q_i$ from $\head(e^i_j)$ to $\tail(e^i_{j+1})$, for $j\in
\{1,\ldots,2n_i\}$, $j$ odd (this path may be edgeless only in the  case where $\head(e^i_j)=\tail(e^i_{j+1})$ is a cut-vertex of ${\rm \Gamma}_{\bf J}$).  We replace $F^i_j$ by a single edge $c^i_j$ (when the corresponding path is edgeless, the edge $c^i_j$ is a loop outside $B$).
We consider the graph $C$ with vertex set $V(P)$ 
and edge set $\{c^i_j\mid
i\in\{1,\ldots,k\},\; E(Q_i)\cap E_R\neq\emptyset,\;
j\in\{1,\ldots,2_{n_i}\},\; j\text{ odd}\}.$

Our strategy consists of a two-step transformation of this embedding. The first 
step creates an embedding of  $C$ inside ${\bf clos}({\bf B})$ without moving the vertices. Indeed, notice that $C$ is embedded in ${\rm \Sigma}_{0}\setminus {\bf B}
$ without crossings such that all the endpoints of the edges in $C$ lie
on the boundary of ${\bf B}.$ As for every odd $j$ in $[2n_i]$, $j+1$ also belongs to $[2n_{i}]$, it follows that none of the endpoints
of $F_{j}^{i}$ can be a terminal.  This, together with the observation above, 
implies that no two edges of $C$ have a common endpoint.

By applying Lemma~\ref{standard} on the $\Sigma_{0}$-embedded outerplanar 
graph $\Gamma=\Gamma_{\bf J}\cup C$ where ${\bf B}$ plays the role of the face $f$,
we obtain a new non-crossing embedding $\Gamma'$.
Notice that $\Gamma'\setminus \Gamma_{\bf J}$
is a new non-crossing embedding of $C$ 
where all of its edges lie inside ${\bf B}.$  
(Recall that none of the  edges of $E_{R}$ is present in this embedding.)
This transformation maps every edge $c_{j}^{i}$ to a new edge inside ${\bf B}$ with the same endpoints.

The second step ``reflects''  the resulting embedding along the axis defined by the path $R$ such that 
each vertex is exchanged with its counterpart.
%
%
 Now define $(c^i_j)'$ so that it connects
  $\tail(e^i_j)$ and $\head(e^i_{j+1})$ -- these are exactly the counterparts of 
$\head(e^i_j)$ and $\tail(e^i_{j+1}).$ Due to symmetry, the $(c^i_j)'$ are
pairwise non-crossing, and none of them crosses a drawing of an edge in $E( P
)\setminus E_R.$  Hence the $(c^i_j)'$ together with the drawing of edges in
$E( P )\setminus E_R$ provide a planar drawing of $ P '$ (where
$(c^i_j)'$ is the drawing of $f^i_j$).  
\imargin{ Why is 
the new $P'$ compatible with $G$'s embedding?}
We finally define (up to isomorphism) the cactus set ${\bf J}'$ of $G$ such that ${\rm \Gamma}_{{\bf J}'}$ is obtained from ${\rm \Gamma}_{\bf J}$ after dissolving the vertices  of $P\cap {\rm \Gamma}_{{\bf J}}$ that are isolated in $P'$.
It is easy to verify that  $(P',{\bf J}')$ is an {\bf F}-patch of $G.$
This concludes the proof of the Claim.\medskip\medskip

Using the Claim above and from Lemma~\ref{lem:colorsequence}, it
follows that $|w|\leq 2^k$, and hence the diameter of $P^*$ is bounded by $2^k.$
 The proof of Theorem~\ref{theo:patchsize} is now complete.
\qed

\imargin{explanation shortened}

\medskip

\medskip
Let $\mathcal L$ be a list of all simple plane
graphs with at 
most $\min\{\ell,\patchsize\}$ edges and no isolated vertices. 
We call a graph in $\mathcal L$ a \emph{completion}. 
As a first step, our algorithm for $\dpc$  computes the list $\mathcal L.$
Obviously, the running time of this process is
bounded by a function depending only  on $k.$

\section{The algorithm for \sc{Planar-Dpc}}
\label{sec:algo}

The fact that the size of ${\mathcal L}$ is bounded by a function of $k$ 
implies that   \dpc\  is in {\sf XP}. Indeed, given the list $\mathcal L$, for
each completion $\tilde{P}\in \mathcal L$ we define the graph $Q_{\tilde{P}}=(V(\tilde{P}),\emptyset)$ and we consider all cactus sets $\tilde{\bf J}$ of $Q_{\tilde{P}}$
where $(\tilde{P},\tilde{\bf J})$ is a $({\rm \Sigma}_{0}\setminus{\bf clos}(\tilde{\bf J}))$-patch of $Q_{\tilde{P}}$ and $V(\tilde{\bf J})=V(\tilde{P}).$  We denote the set of all such pairs $(\tilde{P},\tilde{\bf J})$ by ${\cal J}$ and observe that the number of its elements (up to topological isomorphism of the graph $\tilde{P}\cup {\rm \Gamma}_{\tilde{\bf J}}$) is  bounded by a function of $k.$

For each pair  $(\tilde{P},\tilde{\bf J})\in{\cal J}$, we check whether 
there exists an ${\bf F}$-patch $(P,{\bf J})$  of $G$ such that 
$\tilde{P}\cup {\rm \Gamma}_{\tilde{\bf J}}$ and $P\cup {\rm \Gamma}_{\bf J}$ are 
topologically isomorphic 
and $\DP$ has a solution in the  graph $G\cup P.$ 
As there are $n^{z(k)}$ ways to choose $({P},{\bf J})$
and each check can be done in $O(z_1(k)\cdot n^{3})$ steps,
we conclude  that \dpc\ can be solved in $n^{z_2(k)}$ steps.
In the remainder of the paper, we will prove that the problem is actually in {\sf FPT}.

The main bottleneck is that  there are too many ways
to identify $V(\tilde{\bf J})$ with vertices of  $V({\bf F})$,
because 
we cannot bound $\left| V({\bf F}) \right|$ by a function of $k.$
To overcome this, we associate  with a  positive instance of
$\dpc$ a rooted topological minor $(\tilde{H},\tilde{T})$ of the original graph $G$, that witnesses the fact
that $(\tilde{P},\tilde{\bf J})$ corresponds to the desired ${\bf F}$-patch of $G.$
For convenience, we assume from now on that $\Gamma_{{\bf int}({\Sigma_{0}\setminus {\bf F})}}$ does not contain any loops or multiple edges. In other words, 
none of the open disks corresponding to the connected components 
of $\Sigma_{0}\setminus {\bf F}$ contains less than $3$ vertices in its boundary.
If the input instance violates this property we may enforce it by adding isolated vertices in $G$ 
and modify ${\bf F}$ such that the new vertices belong to its boundary.
Similarly, we restrict  ${\cal J}$ to contain only pairs  $(\tilde{P},\tilde{\bf J})$
where  $\Gamma_{\tilde{\bf J}}$ does not contain any loops or multiple edges.

Given a pair $(\tilde{P},\tilde{\bf J})\in{\cal J}$, we say that a rooted simple graph $(\tilde{H},\tilde{T}=\{a_{1},b_{1},\ldots,$ 
$a_{k},b_{k}\})$ embedded in ${\rm \Sigma}_{0}$, is {\em compatible} with $(\tilde{P},\tilde{\bf J})$ when 

\begin{enumerate}
\item for every $e\in E(\tilde{H})$, $e\subseteq \tilde{\bf J}$,
\item the vertices of $\tilde{H}$ that are also vertices of $\Gamma_{\tilde{\bf J}}$
are  $|V(\tilde{P})|$, while the vertices of  $\tilde{H}$ that are not vertices of $\Gamma_{\tilde{\bf J}}$
are all vertices of $\tilde{T}$,

\item  $V(\tilde{H})\setminus \tilde{T}\subseteq V(\tilde{\bf J})\subseteq V(\tilde{H})$,
\item $\DP(\tilde{P}\cup \tilde{H}, a_{1},b_{1},\ldots,a_{k},b_{k})$ has a solution.
\end{enumerate}

\begin{figure}[t]
\begin{center}
\scalebox{.238}{\includegraphics{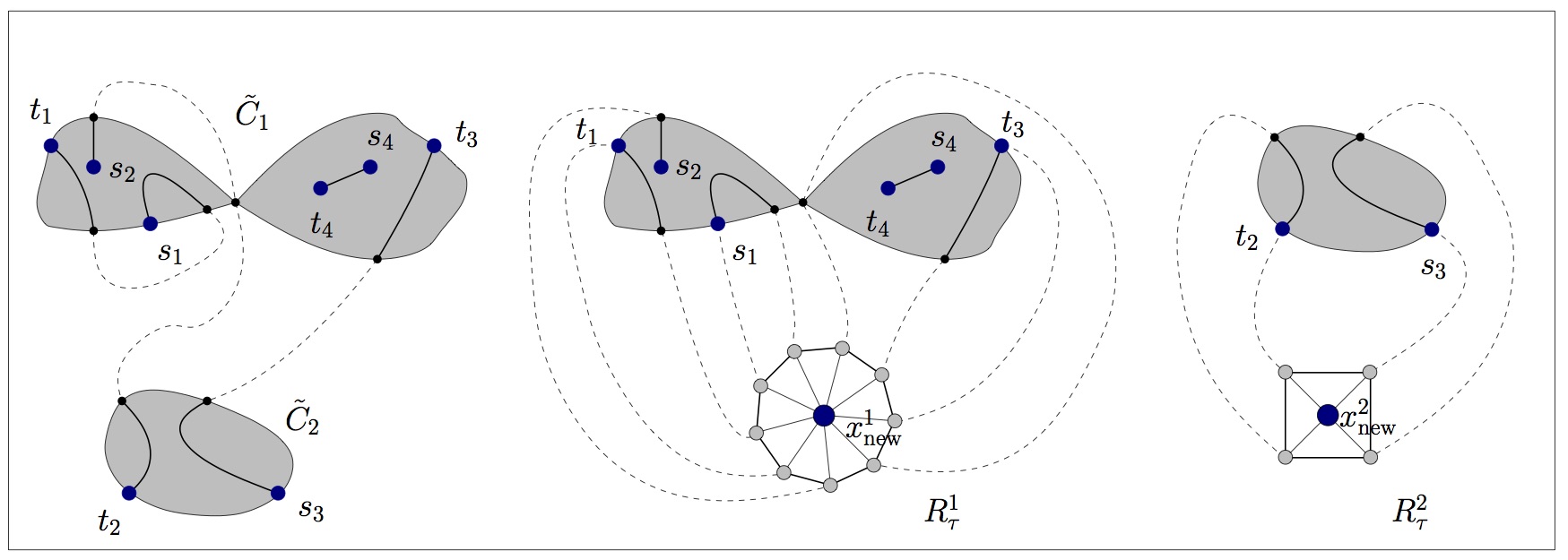}}
\end{center}
\caption{
\small 
In the leftmost image the
dotted lines are the edges of $\tilde{P}.$ Together with the  interior
of the grey areas they form a pair $(\tilde{P},\tilde{\bf J})\in {\cal
  J}.$ $\tilde{\bf J}$ has two weakly connected components. $V(\tilde{\bf
  J})$ consists of the $12$  vertices on the boundary of the grey
areas. 
The solid lines that intersect the open set $\tilde{\bf J}$ are the
edges of the graph $\tilde{H},$ which is compatible with
$(\tilde{P},\tilde{\bf J}).$ Let $\tilde{T} =\{s_1, t_1, \ldots, s_4,
t_4\}.$ The triple $\tau=(\tilde{\bf J},\tilde{H},\tilde{T})\in {\cal H}$
has two  parts. The middle and  the leftmost pictures show how each
of these
parts is enhanced in order to construct the graphs
$R_{\tau}^{1}$ and $R_{\tau}^{2}.$
}
\label{fig:enhancements}
\end{figure}

\noindent We define 
\begin{eqnarray*}
{\cal H} & =& \{(\tilde{\bf J},\tilde{H},\tilde{T})\mid \mbox{ there exists a $(\tilde{P},\tilde{\bf J})\in{\cal J}$ }\\
& &  \mbox{ such that $(\tilde{H},\tilde{T})$ is compatible with $(\tilde{P},\tilde{\bf J})$}
\end{eqnarray*}
and notice that $|{\cal H}|$ is bounded by some function of $k.$
See the leftmost part of Fig.~\ref{fig:enhancements} for an example of a triple in ${\cal H}.$

Assuming that $(P,{\bf J})$ is a solution for $\dpc(G,s_{1},t_{1},\ldots,s_{k},t_{k},\ell,{\bf F})$, consider the parts of the corresponding disjoint paths that lie within
$G.$ The intuition behind the definition above is that $\tilde{H}$ is a certificate 
of these ``partial paths'' in $G.$ Clearly, the number of these certificates is bounded by  $|{\cal H}|$ and they can be
enumerated in $f_{0}(k)$ steps, for some suitable function $f_{0}.$ 
 For example, for the solution depicted in
Fig.~\ref{fig:enhancements}, $\tilde{H}$ consists of $7$ disjoint edges, one
for each subpath within $G.$ 
Our task is to find an {\sf FPT}-algorithm that for every such 
certificate checks whether  the corresponding partial paths 
exist in $G.$

Given an open set ${\bf O}$, a {\em weakly connected component}  of ${\bf O}$ is the interior of some connected component of the set ${\bf clos}({\bf O}).$
Notice that a weakly connected component is not necessarily a connected set.

Let $\bar{\bf F}^{1},\ldots,\bar{\bf F}^{\lambda}$ be the
weakly connected components of the set ${\rm \Sigma}_{0}\setminus {\bf clos}({\bf F}).$
We call such  a component $\bar{\bf F}^{i}$ {\em active} if ${\bf clos}(\bar{\bf F}^{i})\cap T\neq \emptyset.$ We denote  the collection of all active components by ${\cal F}_{\bf F}.$
A crucial observation is that if an ${\bf F}$-patch exists we can always 
replace it by one that bypasses the inactive components.

\begin{lemma}
\label{lemma:fewcc}
\begin{sloppypar}
Let $(G,s_{1},t_{1},\ldots,s_{k},t_{k},{\bf F})$
be an instance for the $\dpc$ problem and let $G'=G[\bigcup_{\bar{\bf F}^{i}\in{\cal F}_{\bf F}}{\bf clos}(\bar{\bf F}^{i})\cap V(G)]$ and ${\bf F}'={\rm \Sigma}_{0}\setminus\bigcup_{\bar{\bf F}^{i}\in{\cal F}_{\bf F}}{\bf clos}(\bar{\bf F}^{i}).$ 
Then 
 $(G',s_{1},t_{1},\ldots,s_{k},t_{k},{\bf F}')$ is an equivalent
 instance.
\end{sloppypar}
\end{lemma}

\paragraph{Proof.}
For the non-trivial direction, we assume that $G$ 
has an ${\bf F}$-patch $(P,{\bf J})$ such that $G\cup P$
contains a collection $\{P_{1},\ldots,P_{k}\}$ paths that certify 
the feasibility  of $\DP(G\cup P, s_{1},t_{1},\ldots,s_{k},t_{k}).$
Let $P'$ be the graph obtained if in
$(\bigcup_{i=1,\ldots,k} P_{i})\cap {\bf clos}({\bf F}')$ we dissolve all vertices 
that are in ${\bf F}'.$ Let also ${\bf J}'$ be the union of all
weakly connected components of the set ${\bf J}$ that contain some open set from  ${\cal F}_{\bf F}.$
Observe that $(P',{\bf J}')$ is an ${\bf F}'$-patch of $G'.$\qed\medskip

By Lemma~\ref{lemma:fewcc}, 
we can assume from now on that the number $\lambda$ of the weakly connected components of the set  ${\rm \Sigma}_{0}\setminus {\bf clos}({\bf F})$ is at most $2k.$ 
Also we {\sl restrict}  ${\cal H}$ so that it contains 
only triples $(\tilde{\bf J},\tilde{H},\tilde{T})$ such that the weakly connected components of the set $\tilde{\bJ}$ are exactly $\lambda.$

\subsection{The enhancement operation} 
\label{sec:enhance}

Consider the triple $\tau=(\tilde{\bf J},\tilde{H},\tilde{T})\in{\cal H}.$  
Let $\tilde{\bJ}^{1},\ldots,\tilde{\bJ}^{\lambda}$ be the weakly connected components of the set $\tilde{\bJ}.$ Then we define 
$\tilde{C}^{i}={\rm \Gamma}_{\tilde{\bJ}^{i}}\cup ({\bf clos}(\tilde{\bJ}^{i})\cap \tilde{H})$ for $i\in\{1,\ldots,\lambda\}$ and we call them {\em parts} of $\tau.$ Also we set $\tilde{T}^{i}=\tilde{T}\cap V(\tilde{C}^{i})$, $1\leq i\leq \lambda.$ We now apply the following {\em enhancement}
operation on each part of $\tau$: 
For $i=1,\ldots,\lambda$,
we consider the sequence ${\cal R}_{\tau}=(R_{\tau}^{1},\ldots,R_{\tau}^{\lambda})$ where $R_{\tau}^{i}$ is the rooted graph $(R'^{i}_{\tau},\tilde{T}^{i}\cup\{x_{\rm new}^{i}\})$ such that $R'^{i}_{\tau}$
is defined as follows.
Take the 
disjoint union of the graph $\tilde{C}^{i}$ and a copy of  the wheel $W_{\mu(\tilde{\bf J}^{i})}$ with center 
$x_{\rm new}^{i}$ and add  $\mu(\tilde{\bf J}^{i})$ edges, called {\em
$i$-external}, between the vertices of $V(\tilde{\bf J}^{i})$ and the peripheral vertices of $W_{\mu(\tilde{\bf J}^{i})}$ such that 
the resulting graph  remains ${\rm \Sigma}_{0}$-embedded and  each vertex
$v\in V(\tilde{\bf J}^{i})$ is incident to $\mu(v)$ non-homotopic
edges not in $\tilde{\bf J}.$ As the graph ${\rm \Gamma}_{\tilde{\bJ}^{i}}$
is connected and planar, the construction of $R'^{i}_{\tau}$ is
possible. Observe also that $R'^{i}_{\tau}\setminus \tilde{\bf J}^{i}$
is {\sl unique up to topological isomorphism.}  To see this, it is enough to verify that for every two vertices in $R'^{i}_{\tau}\setminus \tilde{\bf J}^{i}$ of degree $\geq 3$ there are always 3 disjoint paths connecting them (here we use  our assumption that  $\Gamma_{\tilde{\bf J}}$
does not contain any loops or multiple edges). 

We define ${\cal R}=\{{\cal R}_{\tau}\mid \tau\in{\cal H}\}$
and observe that $|{\cal R}|$ is bounded by a function 
of $k$ because the same holds for $|{\cal H}|.$
 (For an example of the construction of ${\cal R}_{\tau}$, see Fig.~\ref{fig:enhancements})

We now define $(C^{1},\ldots,C^{\lambda})$ 
such that $C^{i}={\rm \Gamma}_{\bar{\bf F}^{i}}\cup ({\bf clos}(\bar{\bf F}^{i})\cap G), i\in\{1,\ldots,\lambda\}.$
We call the graphs in $(C^{1},\ldots,C^{\lambda})$ 
{\em parts} of $G$ and let $T^{i}=T\cap V(C^{i})$, $1\leq i\leq \lambda$, where $T=\{s_{1},t_{1},\ldots,s_{k},t_{k}\}.$  As above we define 
the {\em enhancement} of the parts of $G$ as follows.
For each $i=1,\ldots,\lambda$
we define the rooted graph $G^{*i}=(G^{\prime i},T^{i}\cup \{x^{*i}_{{\rm new}}\})$  where $G^{\prime i}$  is defined as follows:
 take the 
disjoint union of $C^{i}$ and the wheel $W_{\mu(\bar{\bf F}^{i}) }^{*}$ with center 
$x_{\rm new}^{*i}$ and add  $\mu(\bar{\bf F}^{i})$ edges, called {\em
$*i$-external,}  between the vertices of $V(\bar{\bf F}^{i})$ and the peripheral vertices of $W_{\mu(\bar{\bf F}^{i})}^{*}$ such that 
the resulting graph remains ${\rm \Sigma}_{0}$-embedded and each vertex $v\in V(\bar{\bf F}^{i})$ is incident to $\mu(v)$ non-homotopic edges. As above, each $G^{\prime i}$ is possible to construct and  $G^{\prime i}\setminus \bar{\bf F}^{i}$ is unique up to topological isomorphism.

The purpose of the above definitions is twofold. First, 
they help  us to treat separately each of the parts of 
$G$ and try to match them with the correct 
parts of $\tau.$ Second, the addition of the wheels
to each part gives rise to a single, uniquely embeddable interface, between the part and its ``exterior''
and this helps us to treat embeddings as abstract graphs.
Therefore, to check whether a part of $\tau$
is realizable within the corresponding part of $G$, we can use the rooted version of the topological minor relation on graphs as defined in Section~\ref{sec:prel}.

\subsection{The stretching lemma}
\label{sec:stretch} 

A bijection $\rho$ from $\tilde{T}=\{a_{1},b_{1},\ldots,a_{k},b_{k}\}$ to $T=\{s_{1},t_{1},\ldots,s_{k},t_{k}\}$ 
is {\em legal} if for every  $i\in\{1,\ldots,k\}$, 
there exists some  $j\in\{1,\ldots,k\}$ such that
$\rho((a_i,b_{i}))=(s_j,t_{j}).$

Let $\tau\in{\cal H}$ and let $\rho$ be a legal bijection from 
$\tilde{T}$ to $T$ and 
let $\rho_{i}$ be the restriction of $\rho$ in $\tilde{T}_{i}.$ We say that ${\cal R}_{\tau}=(R_{\tau}^{1},\ldots,R_{\tau}^{\lambda})$
is {\em $\rho$-realizable} in $G$ if there exists 
a bijection $\phi \colon  \{1,\dots, \lambda\}\rightarrow \{1,\ldots,\lambda\}$ such that for $i=1,\ldots,\lambda$,
$R_{\tau}^{i}$ is a $\hat{\rho_{i}}$-rooted topological minor of $G^{*\phi(i)}$ were $\hat{\rho}_{i}=\rho_{i}\cup\{(x_{\rm new}^{i},x_{\rm new}^{*\phi(i)})\}.$

By enumerating all possible bijections $\phi$,
we enumerate all possible correspondences between the parts of  of $G$
and the parts of $\tau.$ In order to simplify notation, we
assume in the remainder of this section that $\phi$ is the identity function. 

The following lemma is crucial. It shows that when  $R_{\tau}^{i}$ is
a topological minor of $G^{*i}$ we can always assume  that all vertices and edges of $\tilde{C}^{i}$ are mapped via $\psi_{0}$ and $\psi_{1}$
to vertices and paths in  ${\bf clos}(\bar{\bf F}^{i})$; 
the wheel $W_{\mu(\tilde{\bJ}^{i})}$ 
is mapped to a ``sub-wheel'' of  $W_{\mu(\bar{\bf F}^{i})}$
while $i$-external edges are mapped to $*i$-external edges. 
This will be useful in the
proof of Lemma~\ref{moster}, as the  $i$-external edges represent the
interface of  the completion $\tilde{P}$ with $\tilde{C}^i.$ The
topological minor relation certifies that the  same interface is feasible  
between  the corresponding part ${C}^i$ of $G$ and its ``exterior''. 
Lemma~\ref{lem:helio}  establishes also that the image of 
${\rm \Gamma}_{\tilde{\bJ}^{i}}$ can be ``stretched'' so that it falls on   ${\rm \Gamma}_{\bar{\bf F}^{i}}.$ As
all the  vertices in $V(\bar{\bf F}^{i})$ are within distance $2$ from
the artificial terminal $x_{\rm new}^{*i}$ in $G^{*i},$ this allows us later
in the proof of Lemma~\ref{irelmos} to locate within $G^{*i}$ the possible
images of $V(R_{\tau}^{\prime i})$ in a neighborhood of the
terminals. It is then safe to look for an irrelevant vertex
``far away'' from this neighborhood. 

\begin{lemma}
\label{lem:helio}
Let $R_{\tau}^{i}$ be a $\hat{\rho_{i}}$-rooted topological minor of $G^{*i}$ where $\hat{\rho}_{i}=\rho_{i}\cup\{(x_{\rm new}^{i},x_{\rm new}^{*i})\}$, for $i=1,\ldots,\lambda.$ Let also $\psi_{0}^{i}$ and $\psi_{1}^{i}$ be the functions (cf.  Section~\ref{sec:prel}) certifying this topological minor relation.  Then $\psi_{0}^{i}$ and $\psi_{1}^{i}$ can be modified so that the following properties 
are satisfied.
\begin{enumerate}
\item if $\tilde{e}$ is an edge of the wheel $W_{\mu(\tilde{\bJ}^{i})}$ incident to $x_{\rm new}^{i}$,
then $\psi_{1}^{i}(\tilde{e})$ is an edge incident to $x_{\rm new}^{*i}.$          
\item if $\tilde{e}$ is an edge of the wheel $W_{\mu(\tilde{\bJ}^{i})}$ not incident to $x_{\rm new}^{i}$,
then $\psi_{1}^{i}(\tilde{e})$ is an $x_{\rm new}^{*i}$-avoiding path of  $W_{\mu(\bar{\bf F}^{i})}^{*}.$ 
Moreover, each edge in $W_{\mu(\bar{\bf F}^{i})}^{*}$ is contained in a path $\psi_{1}^{i}(\tilde{f})$
for some edge $\tilde{f}$ of the cycle $W_{\mu(\tilde{\bf J}^{i})}^{*}\setminus x_{\rm new}^{i}.$ 
%
\item if $\tilde{e}$ is an $i$-external edge between
$V(\tilde{\bJ}^{i})$ and $V(W_{\mu(\tilde{\bJ}^{i})})\setminus\{x_{\rm
new}^{i}\}$, then $\psi_{1}(e)$ is a path consisting of an    
$*i$-external edge between $V(\bar{\bf F}^{i})$ and $V(W_{\mu(\bar{\bf F}^{i})})\setminus\{x_{\rm new}^{*i}\}.$
\item $\psi_{0}^{i}(V(\tilde{\bJ}^{i}))\subseteq V(\bar{\bf F}^{i}).$   
\end{enumerate}

\end{lemma}

\paragraph{Proof.}
Our {\sl first step} is to enforce Properties~1~and~2. For each edge $\tilde{e}=\{x_{\rm new}^{i},x\}$ of $R^{i}_{\tau}$, let $\tilde{e}'=\{x,x'\}$ be the unique edge of $R^{i}_{\tau}$ that is not an edge of $W_{\mu(\tilde{\bJ}^{i})}.$
Consider the path $P_{\tilde{e}}$ that is formed by the concatenation of $\psi_{1}^{i}(\tilde{e}')$ and $\psi_{1}^{i}(\tilde{e})$
and let $\hat{x}$ be the neighbor of $\psi_{0}^{i}(x_{\rm new}^{i})=x_{\rm new}^{*i}$ in the path $\psi_{1}^{i}(P_{\tilde{e}}).$ For each such $\tilde{e}$ and its incident vertex $x,$ we simultaneously update 
$\psi_{0}^{i}$ such that $\psi_{0}^{i}(x)=\hat{x}.$ Accordingly, for every $\tilde{e}=\{x_{\rm new}^{i},x\}$ we simultaneously update 
$\psi_{1}^{i}$ so that $\psi_{1}^{i}(\{x_{\rm new}^{i},x\})$ is the path consisting of the edge $\{x_{\rm new}^{*i},\hat{x}\}$ (enforcing Property~1) and $\psi_{1}^{i}(x,x')$ is the 
subpath of $P_{\tilde{e}}$ between $\hat{x}$ and $\psi_{0}^{i}(x').$ We also update $\psi_{1}^{i}$ such that 
for each $\tilde{e}_{1}=\{x_{\rm new}^{i},x_{1}\}$ and $\tilde{e}_{2}=\{x_{\rm new}^{i},x_{2}\}$ where $x_{1}$ and $x_{2}$ are adjacent,
we set $\psi_{1}^{i}(\{x_{1},x_{2}\})$ to be the $(\hat{x}_{1},\hat{x}_{2})$-path of $W_{\mu(\bar{\bf F}^{i})}^{*}$ that does not meet any other vertex of $\psi_{0}^{i}(V(W_{\mu(\tilde{\bJ}^{i})}))$, enforcing Property~2.

So far, we have enforced that  for each $i\in\{1,\ldots,\lambda\}$, the wheel $W_{\mu(\bar{\bf F}^{i})}$ is mapped via $\psi_{0}^{i}$ and $\psi_{1}^{i}$ to the wheel $W_{\mu(\tilde{\bJ}^{i})}$ (by slightly abusing the notation, we can say that $\psi_{1}(W_{\mu(\tilde{\bJ}^{i})})=W_{\mu(\bar{\bf F}^{i})}$).
Notice that Properties 1 and 2 imply that 
all vertices and edges of $\tilde{C}^{i}$ are mapped via $\psi_{0}$ and $\psi_{1}$
to vertices and paths of  ${\bf clos}(\bar{\bf F}^{i}).$ 

Our {\sl second step} is to enforce Properties 3 and 4. 
The transformation that we describe next, essentially
``stretches'' the image of $\tilde{\bJ}^{i}$ until it hits 
from within the boundary of $\bar{\bf F}^{i}.$
Let $\tilde{e}=\{x,x'\}$ be an edge of ${R}^{i}_{\tau}$
where  $x\in V(\tilde{\bJ}^{i})$ and $x'\in V(W_{\mu(\tilde{\bJ}^{i})})\setminus\{x_{\rm new}^{i}\}.$ We also set $P_{\tilde{e}}=\psi_{1}^{i}(\tilde{e}).$ By definition,
$P_{\tilde{e}}$ is a path in $G^{*i}$ which 
starts from $\psi_{0}^{i}(x)$ and ends at $\psi_{0}^{i}(x').$
As $\psi_{0}^{i}(x)$ is a point of ${\bf clos}(\bar{\bf F}^{i})$ and 
$\psi_{0}^{i}(x')$ is not, we can define $\hat{x}$ as the first  
vertex of $P_{\tilde{e}}$ that  is a vertex of $V(\bar{\bf F}^{i}).$
For each such $\tilde{e}$ and its incident vertex $x$,
we simultaneously update 
$\psi_{0}^{i}$ such that $\psi_{0}^{i}(x)=\hat{x}.$ 
Accordingly, for every $\tilde{e}=\{x,x'\}$ we simultaneously update 
$\psi_{1}^{i}$ so that $\psi_{1}^{i}(\{x,x'\})$ is the path consisting of the single edge $\{\hat{x},\psi_{0}^{i}(x')\}$ -- thus enforcing Property 3 --  and for each edge $\{x_{1},x_{2}\}$ of ${\rm \Gamma}_{\tilde{\bf J}^{i}}$ we update $\psi_{1}^{i}$
so that $\psi_{1}^{i}(\{x_{1},x_{2}\})$ is a $(\hat{x}_{1},\hat{x}_{2})$-path avoiding any other vertex of $\psi_{0}^{i}(V(\tilde{\bf J}^{i})).$ As now all images of the vertices 
and the edges of ${\rm \Gamma}_{\tilde{\bf J}^{i}}$ lie on ${\rm \Gamma}_{\bar{\bf F}^{i}}$, Property 4 holds. \qed\medskip

\subsection{Reducing \dpc\ to  topological minor testing}
\label{sec:reduce}

\begin{lemma} 
\label{moster}
$\dpc(G,t_1,s_1,\ldots,t_k,s_k,\ell,{\bf F})$ has a solution if and only if
there exists a $\tau=(\tilde{\bf J},\tilde{H},\tilde{T})\in{\cal H}$
and a legal bijection $\rho \colon  \tilde{T}\rightarrow T$  such that ${\cal R}_{\tau}$ is $\rho$-realizable in $G.$
\end{lemma}

%
%

\paragraph{Proof.}
Suppose  that $(P,{\bf J})$ is a solution for $\dpc(G,t_1,s_1,\ldots,t_k,s_k,\ell,{\bf F})$ giving rise to a collection
${\cal P}=\{P_{1},\ldots,P_{k}\}$ of disjoint paths in $G\cup P$ where 
$P_{i}$ connects $s_{i}$ with $t_{i}.$

From Theorem~\ref{theo:patchsize}, we can assume that 
$|V({\bf J})|$ is bounded by a suitable function of $k$ and therefore, $(P,{\bf J})$ is isomorphic 
to a member $(\tilde{P},\tilde{\bf J})$ of ${\cal J}$, in the sense 
that $\tilde{P}\cup {\rm \Gamma}_{\tilde{\bf J}}$ and $P\cup {\rm \Gamma}_{\bf J}$ are 
topologically isomorphic. 
Let $(H,T)$ be the rooted subgraph of $G$ formed by the edges of the paths in ${\cal P}$ that are also edges of $G.$ We now define the rooted graph $({H}',T)$
by dissolving all non-terminal vertices in the interior of ${\bf J}$
and observe that  $({H}',T)$ is isomorphic, with respect to some bijection $\omega$,  to 
some $(\tilde{H},\tilde{T})$ that is compatible with $(\tilde{P},\tilde{\bf J}).$
Therefore $\tau=(\tilde{\bf J},\tilde{H},\tilde{T})\in{\cal H}$ and 
let $\rho$ be the correspondence between $\tilde{T}$ and $T$,  
induced by $\omega.$ By partitioning $\omega$ with respect
to the weakly connected components of the set $\tilde{\bJ}$, we generate  
a bijection $\phi$ between the parts $\{\tilde{C}^{1},\ldots,\tilde{C}^{\lambda}\}$ of $\tau$ and the parts $\{{C}^{1},\ldots,{C}^{\lambda}\}$ of $G$ such that a subdivision of $\tilde{C}^{i}$ is topologically isomorphic to a subgraph of $C^{\phi(i)}$, $i\in\{1,\ldots,\lambda\}.$
This topological isomorphism can be extended in the obvious manner to 
the graphs obtained after the enhancement operation applied 
to the parts of $\tau$ and the parts of $G.$ This translates to the fact that each of the resulting  $R_{\tau}^{i}$ is a $\rho_{i}$-rooted topological minor of $G^{*\phi(i)}$ where each $\rho_{i}$ is obtained by the restriction of $\rho$ to the vertices of $V(R_{\tau}^{i})$ plus the pair $(x_{\rm new}^{i},x_{\rm new}^{*i}).$ We conclude that ${\cal R}_{\tau}$ is $\rho$-realizable in $G.$\medskip

For the opposite direction, let $\tau\in{\cal H}.$  By renumbering if
necessary the parts of $G,$ let $R_{\tau}^{i}$ be a $\hat{\rho_{i}}$-rooted topological minor of $G^{*i}$ were $\hat{\rho}_{i}=\rho_{i}\cup\{(x_{\rm new}^{i},x_{\rm new}^{*i})\}$, for $i=1,\ldots,\lambda.$ Let also $\psi_{0}^{i}$ and $\psi_{1}^{i}$ be the functions satisfying the four properties of  Lemma~\ref{lem:helio}.

\begin{figure}[h]
\begin{center}
\scalebox{.31973}{\includegraphics{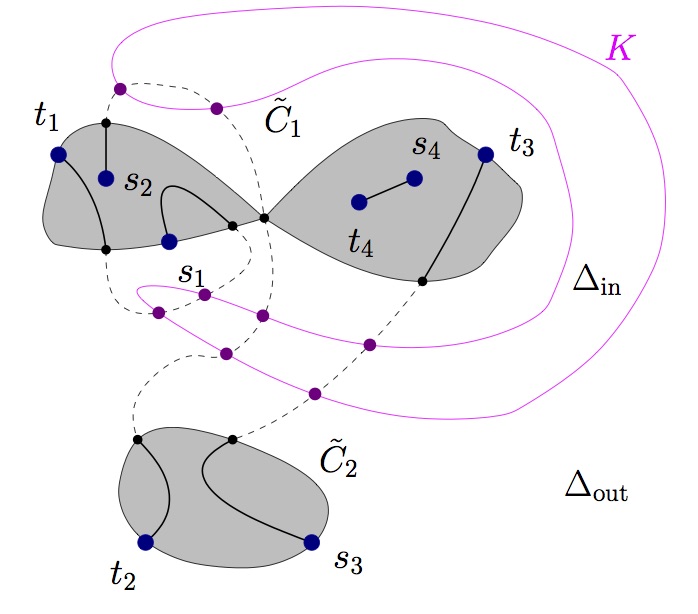}}
\end{center}
\caption{The graph $\tilde{H}_{K}^{+}$ when $\tilde{H}^{+}$ is the leftmost
graph of Figure~\ref{fig:enhancements}.}
\label{enhmore}
\end{figure}

For the given $\tau=(\tilde{\bf J},\tilde{H},\tilde{T})\in {\cal H}$ we assume w.l.o.g. that $|V(\tilde{\bf J})|$ is minimal. Recall that there exists 
a $\tilde{P}$ such that $(\tilde{H},\tilde{T})$ is compatible 
with $(\tilde{P},\tilde{\bf J}).$ Let $\tilde{H}^{+}=\tilde{H}\cup{\rm \Gamma}_{\tilde{\bf{J}}}\cup \tilde{P}.$ From Lemma~\ref{topolemma}, there exists a curve $K$ where $K\subseteq {\rm \Sigma}_{0}\setminus {\bf clos}(\tilde{{\bf J}})$ and such that 
$K$ intersects each edge of $\tilde{P}$ twice. Let $\Delta_{\rm in}$ and $\Delta_{\rm out}$ be the two connected components of the set  ${\rm \Sigma}_{0}\setminus K$ and w.l.o.g. we may assume that ${\bf clos}(\tilde{\bf J})\subseteq \Delta_{\rm out}.$ If we consider all connected components 
of the set $K\setminus \tilde{H}^{+}$ as edges and take their 
union with the graph obtained by $\tilde{H}^{+}$ after  subdividing the edges of $\tilde{P}$ at the points of $K\cap \tilde{H}^{+}$, we create a new graph $\tilde{H}^{+}_{K}$ (see Figure~\ref{enhmore} for an example).
We also define $\tilde{H}^{+}_{K,{\rm in}}=\tilde{H}^{+}_{K}\cap \Delta_{\rm in}$ and $\tilde{H}^{+}_{K,{\rm out}}=\tilde{H}^{+}_{K}\cap {\bf clos}(\Delta_{\rm out}).$
For $i\in\{1,\ldots,\lambda\}$, we define 
$\tilde{H}^{+(i)}_{K,{\rm out}}$ as follows:
first take the graph obtained by 
$\tilde{H}^{+}_{K,{\rm out}}$ if we remove all edges except 
those that have at least one endpoint in $\tilde{C}^{i}$
and those that belong in $K$ and, second, dissolve in this 
graph all vertices that have degree 2. If we now add 
a new vertex $x_{\rm new}^{i}\in \Delta_{\rm in}$ and make it adjacent 
to all remaining vertices in $K$, it follows, by the minimality of the choice of $\tau$, that the resulting graph is 
isomorphic to $R_{\tau}^{i}.$
In the above construction, we considered the vertex set 
$Q=K\cap \tilde{H}^{+}\subseteq V(H^{+}_{K})$ 
and we slightly abuse notation by denoting by $Q$ 
also its counterpart in $R_{\tau}^{i}.$
We also 
denote by $\pi_{Q}$ the cyclic ordering of $Q$ defined by the order of appearance of the vertices in $Q$ along $K.$
Similarly, for $i\in\{1,\ldots,\lambda\}$, we set $Q_{i}=Q\cap V(\tilde{H}_{K,{\rm out}}^{+(i)})$ and we denote by $\pi_{Q_{i}}$ the induced sub-ordering of $\pi_{Q}.$
We denote by  $\tilde{E}^{i}=\{\tilde{e}_{1}^{i},\ldots,\tilde{e}_{\mu(\tilde{\bf J}^{i})}^{i},\tilde{e}_{1}^{i}\}$  the edges of $R_{\tau}^{i}$, not in $W_{\mu({\tilde{\bJ}^{i}})}$, that are incident to the vertices of $Q_{i}$, cyclically ordered according to $\pi_{Q_{i}}.$

We now take the graph $G^{*i}$ and remove from it 
all edges outside  ${\bf clos}(\bar{\bf F}^{i})$ except from the images 
of the edges of $\tilde{E}^{i}$ (observe that ${\bf clos}(\bar{\bf F}^{i})$ is 
always a connected set while this is not necessarily the case for $\bar{\bf F}^{i}$). 
From Property~3 in Lemma~\ref{lem:helio}, these 
images are just edges between  $V(\bar{\bf F}^{i})$ and $V(W_{\mu(\bar{\bf F}^{i})})\setminus\{x_{\rm new}^{*i}\}.$
We denote the resulting graph by $G^{*i-}$ and observe that it can be drawn  in ${\rm \Sigma}_{0}$ in a way 
that the vertices in $\psi_{0}(Q_{i})$ lie on a virtual 
closed curve in accordance with the cyclic ordering 
$\pi_{Q_{i}}$ of their $\psi_{0}^{i}$-preimages.
We consider the disjoint union $G^{*}_{\rm out}$ of all $G^{*i-}$
and we observe that it is possible to embed it
in ${\rm \Sigma}_{0}$ such that all vertices in $Q^{*}=\bigcup_{i\in\{1,\ldots,\lambda\}}\psi_{0}^{i}(Q_{i})$ lie on the same closed curve $K^{*}.$ By the way each $\pi_{Q_{i}}$ is defined 
from $\pi_{Q}$ and the bijection between $Q$ and $Q^{*}$ we directly obtain that 
the following properties are satisfied:
(i) the new embedding is planar,
(ii) $G^{*}_{\rm out}$ is a subset of one of the connected components of the set ${\rm \Sigma}_{0}\cap K^{*}$,
(iii) for all $i\in\{1,\ldots,\lambda\},$ the cyclic ordering of $\psi_{0}^{i}(Q_{i})$ is an induced sub-ordering of  
the cyclic ordering defined by the way the vertices of
$Q^{*}$
are being arranged along $K^{*}.$
Note that $\psi_{0}=\bigcup_{i\in\{1,\ldots,\lambda\}}\psi_{0}^{i}$ defines a bijection from the vertices 
of $Q$ to the vertices of $Q^{*}.$
Our next step is to obtain the graph  $G^{*}_{\rm in}$
by taking $\tilde{H}^{+}_{K,{\rm in}}$
and renaming each vertex $x\in Q$ to $\psi_{0}(x).$
Now the graph $G^{+}=G^{*}_{\rm in}\cup G^{*}_{\rm out}$
consists of the original graph $G$ and a collection ${\cal Z}$ of 
paths of length 3. 

We are now ready to define the desired 
${\bf F}$-patch $(P,{\bf J})$ of $G.$ 
The edges of $P$ are created  by dissolving
all internal vertices  of each path in ${\cal Z}.$
Also, ${\bf J}$ can be any cactus set for which
the embedding of ${\rm \Gamma}_{\bf J}$ 
results from the embedding of ${\rm \Gamma}_{{\bf F}}$
after dissolving all vertices of $V({\bf F})$ that 
are not in $\psi_{0}(V(\tilde{\bf J})).$
\qed\medskip

In~\cite{KawarabayashiW2010asho}, Grohe, Kawarabayashi, Marx, and Wollan gave an 
$h_{1}(k)\cdot n^{3}$ algorithm for checking rooted topological minor testing, where $h_{1}$ is some  computable function.  
Combining their algorithm with Lemma~\ref{moster},  we obtain an $h_{2}(k)\cdot n^{3}$ algorithm for 
\dpc\ (here, again, $h_{2}$ is some computable function). Therefore, $\dpc\in {\sf FPT}.$ 

In the next section, we show how to obtain the improved running 
time claimed in Theorem~\ref{theo:main}.

\section{Applying the irrelevant vertex technique}
\label{sec:irrel}

Lemma~\ref{moster} established that for every candidate patch
$\tilde{P}$, certifying  its feasibility for \dpc\ on $G$ reduces to finding a rooted
topological minor.  By the results of~\cite{GroheKMW10find}, \dpc\ 
is in {\sf FPT} by an $h_{2}(k)\cdot n^{3}$ algorithm. In this section we show that this
running time can be improved. We first explain the intuition behind this
improvement.

%
%

\subsection{Proof strategy}

At a high level, we have  reduced the  validity of $\tilde{P}$
to the problem of determining whether 
a $q$-vertex rooted graph $H$ is a rooted topological 
minor of an $n$-vertex graph $G$, where $q$ is bounded by some function of $k.$
By Lemma~\ref{lem:helio},  we can assume that
the images of the vertices of $H$ are either terminals of $G$
or lie on the boundary of the same face of $G$.
%
%
This observation makes it possible to directly 
employ, in Lemma~\ref{irelmos}, the {\sl irrelevant vertex technique}~\cite{RobertsonS-XIII}.

In particular,  if the treewidth of $G$ 
is big enough, one can  detect a sufficiently large 
set of concentric cycles that are away from the images of the vertices of $H$ in $G$;
this is possible due to Lemma~\ref{lem:helio}. 
Using the ``vital linkage'' theorem of Robertson and 
Seymour~\cite{RobertsonS-XXI,RobertsonS-XXII} (see also~\cite{KawarabayashiW2010asho}),  we obtain
that the topological minor  mapping can be updated so that 
the  realization of $H$ avoids 
the inner
cycle of this collection. Therefore, the removal 
of any of the vertices of this cycle creates an equivalent instance of
the problem with a smaller number of  vertices.
By repeating this vertex-removal operation, we end up
with a graph whose treewidth is bounded by some function of $q.$ In
this case, 
since the rooted variant of the topological minor checking problem is
definable in Monadic Second Order Logic (\mso)  (see
Lemma~\ref{lemma:msol} in the Appendix),  the problem 
can be solved in a linear number of steps according to Courcelle's Theorem. 

For the running time of our algorithm, we use the fact  that the detection of an irrelevant 
vertex in planar graphs requires to find a vertex that is ``far enough'' 
from all the terminals. As this can be done by standard BFS in $O(n)$ steps
and at most $n$ such vertices are deleted, the overall complexity of the 
algorithm is $f(k)\cdot n^{2}.$

%

\subsection{Treewidth and linkages}

Let $G$ be a graph. A {\em linkage} in $G$ is a set of pairwise disjoint paths of it. The {\em endpoints} of a linkage ${\cal L}$ are the endpoints  of the paths in ${\cal L}.$  The {\em pattern} 
of ${\cal L}$ is defined as 
$$\pi({\cal L})=\{\{s,t\}\mid {\cal L} \mbox{~contains a path from $s$ to $t$}\}$$

Consider now the rooted graph $G=(G',T)$ where $G'$ is a ${\rm \Sigma}_{0}$-embedded  graph. We call a cycle of $G'$ {\em $T$-respectful} if all the vertices of $T$ are inside one of the two connected components of the set ${\rm \Sigma}_{0}\setminus C.$
Given a $T$-respectful cycle $C$ of $G'$ 
we denote by  $\Delta_{\rm ext}(C)$ the connected component of  the set ${\rm \Sigma}_{0}\setminus C$
that contains $T$ and by $\Delta_{\rm int}(C)$ the other.
A sequence $C_{1},\ldots,C_{k}$ of $T$-respectful cycles in $G$ is {\em $T$-concentric}
if $\Delta_{\rm ext}(C_{1})\subseteq\cdots\subseteq \Delta_{\rm ext}(C_{k}).$

\begin{figure}[ht]
\begin{center}
\scalebox{0.21965}{\includegraphics{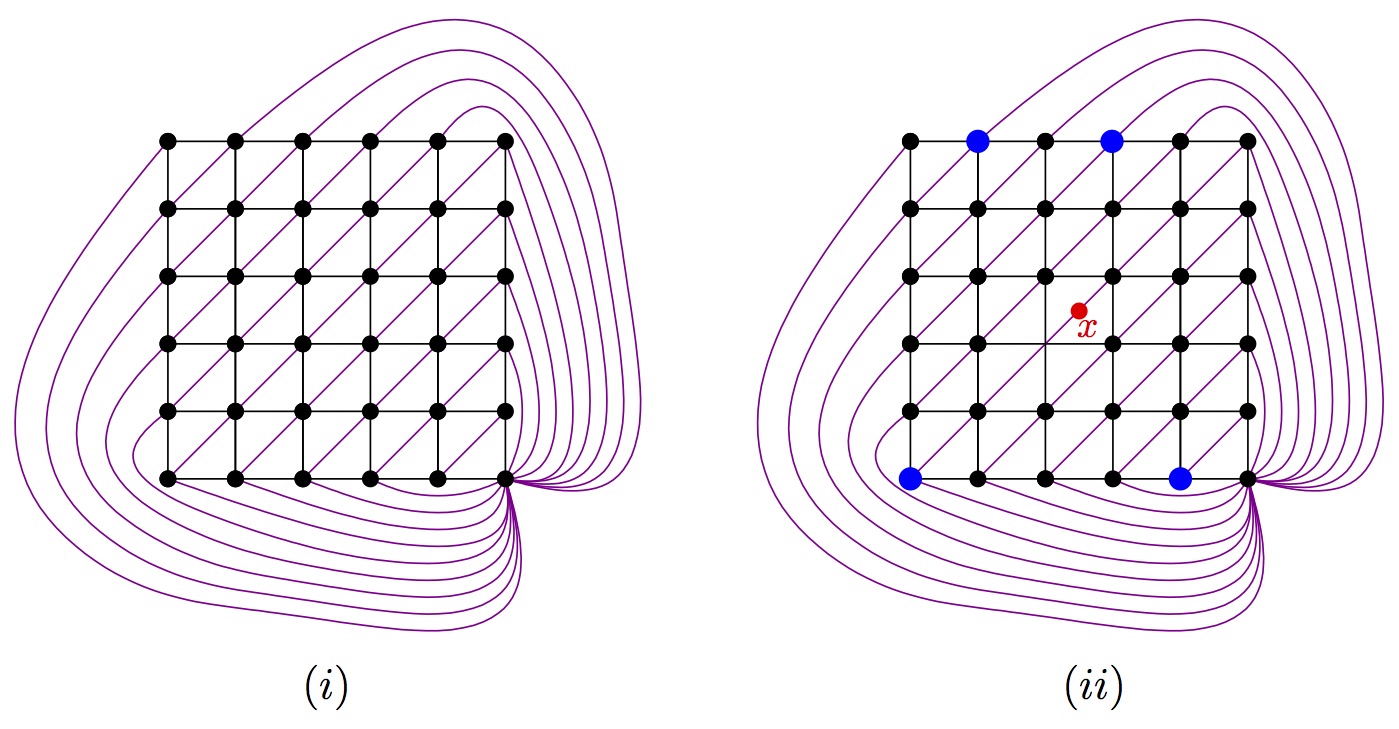}}
\end{center}
\caption{The leftmost graph is ${\rm \Gamma}_{6}$. The insulation of $x$ in the rightmost graph is equal to $2$ (the terminals are the blue vertices).}
\label{fig-gamma-k}
\end{figure}

Let ${\rm \Gamma}_{k}$  ($k\geq 2$) be the graph obtained from the  $(k\times k)$-grid by
triangulating internal faces of the $(k\times k)$-grid such that all internal vertices become  of degree $6$,
all non-corner external vertices are of degree 4,
and then one corner of degree two is joined by edges -- we call them {\em external} -- with all vertices
of the external face (the {\em corners} are the vertices that in the underlying grid have  degree two).
Graph ${\rm \Gamma}_6$ is shown in Fig.~\ref{fig-gamma-k}.$(i)$. 
We also define the graph ${\rm \Gamma}_{k}^{*}$ as the graph obtained from ${\rm \Gamma}_{k}$ if we remove all its external edges.
In ${\rm \Gamma}_{k}^{*}$ we call all vertices incident to its unique non-triangle face {\em perimetric}.
%
%
%

Let $G=(G',T)$ be a rooted ${\rm \Sigma}_{0}$-embedded 
graph. Given $x\in V(G')$ we define the {\em insulation between $x$ and $T$}, denoted ${\bf ins}_{T}(x)$ 
as the maximum length of a sequence of $T$-concentric
cycles $C_{1},\ldots,C_{l}$ in $G$ such that (i) $\{x\}\cap \bigcup_{i\in\{1,\ldots, l\}}C_{i}=\emptyset$ and (ii) every line of ${\rm \Sigma}_{0}$ connecting $x$ and some vertex of $T$ meets every $C_{i}, i\in\{1,\ldots,l\}$ For an example, see Figure~\ref{fig-gamma-k}.$(ii)$.
We define the {\em planar thickness} of a rooted graph $G=(G',T)$ as follows:
$${\bf pth}_{T}(G)=\max \{{\bf ins}_{T}(x)\mid x\in V(G')\}.$$
 It is easy to verify that {\bf pth} is closed under contractions. In other words, the following holds.
\begin{lemma}
\label{cutoc}
Let $G_{1}=(G_{1}',T_{1})$ and $G_{2}=(G_{2}',T_{2})$ be two rooted graphs where $G_{1}'$ is a $\sigma$-contraction of $G_{2}'$ and $T_{1}=\sigma(T_{2}).$ Then ${\bf pth}_{T_{1}}(G_{1})\leq {\bf pth}_{T_{2}}(G_{2}).$
\end{lemma}

According to the following lemma, if the treewidth of $G$ is big enough
then every constant radius ball around $T$ is sufficiently insulated from some vertex of $G.$

\begin{lemma}
\label{apost}
Let $G=(G',T)$ be a rooted plane graph where 
$\tw(G')\geq 24\cdot (2l+2r+2)(\lceil \sqrt{|T|+1}\rceil)+49.$ Let also $T'=\bigcup_{t\in T}N^{r}_{G'}(t).$
Then ${\bf pth}_{T'}(G)\geq l.$
\end{lemma}

\proof{By using~\cite[Lemma 6]{FominGT09cont}, the graph $G'$ contains as a $\sigma$-contraction the
graph $H'={\rm \Gamma}_{(2l+2r+2)(\lceil \sqrt{|T|+1}\rceil)}.$ 
Define the rooted graph $H=(H', T_{H})$ where 
$T_{H}=\sigma(T).$ We consider a vertex packing of $H'$ in
$|T|+1$ copies of ${\rm \Gamma}_{2l+2r+2}^{*}.$ By the pigeonhole principle and the fact that $|T_{H}|\leq |T|$, one, say $Z$, of these copies does 
not contain any vertex in $T_{H}.$
By Observation~\ref{obs:distsl}, contractions do not increase distances, hence $\sigma(T') \subseteq N_{H'}^{r}(T_{H}).$
Let $Z'$ be the subgraph of $Z$ that is isomorphic to ${\rm \Gamma}^{*}_{2l+2}$ 
whose vertices have distance at least $r$ from the perimetric 
vertices of $Z$ and therefore are also at distance strictly greater than $r$ from $T_{H}.$
We conclude that $V(Z')\cap \sigma(T')=\emptyset.$
Notice that $Z'$ contains $l+1$ $\sigma(T')$-concentric cycles $C_{1},\ldots,C_{l},C_{l+1}$ that are also $\sigma(T')$-concentric cycles of  $H.$ 

Let $x\in C_{l+1}.$ Then  the cycles $C_{1},\ldots,C_{l}$ certify that
${\bf ins}_{\sigma(T')}(x)\geq l.$ Therefore ${\bf pth}_{\sigma(T')}(H')\geq l.$ By Lemma~\ref{cutoc}, we conclude that ${\bf pth}_{T'}(G)\geq l.$\qed
}

We need the following theorem from~\cite{RobertsonS-XXII}. We present it using the terminology of~\cite{DawarGK07loca}.

 \begin{proposition}
\label{seymlink}
There is a computable function $g$ such that the following holds:
Let ${\rm \Gamma}$ be a ${\rm \Sigma}_{0}$-embedded plane graph, ${\cal L}$ be a linkage of ${\rm \Gamma}$ and let $T$ be the set of vertices in the pairs of  $\pi({\cal L}).$
Let also $C_{1},\ldots,C_{g(|\pi({\cal L})|)}$ be $T$-concentric  cycles of ${\rm \Gamma}.$ Then there is a linkage ${\cal L}'$ with the same pattern as ${\cal L}$ such that all paths in ${\cal L}$ are contained in $\Delta_{\rm ext}(C_{g(|\pi({\cal L})|)}).$
 \end{proposition}

\subsection{The algorithm}

Using Lemmata~\ref{lem:helio},~\ref{cutoc},~\ref{apost}, and Proposition~\ref{seymlink}, we can prove the following result.

\begin{lemma} 
\label{irelmos}
There is an \fpt algorithm, running in $f_{2}(k)\cdot n^{2},$ for some
function $f_{2}$, that given a $\tau=(\tilde{\bf
  J},\tilde{H},\tilde{T})\in{\cal H}$ and a legal bijection $\rho
\colon  \tilde{T}\rightarrow T,$  checks whether ${\cal R}_{\tau}$ is $\rho$-realizable in $G.$
\end{lemma}

\proof{As the number of bijections $\phi \colon \{1,\dots, \lambda\}\rightarrow \{1,\ldots,\lambda\}$
is bounded by a function of $k$ it is enough to show how 
to check in {\sf FPT} time
whether,  for $i=1,\ldots,\lambda$,
$R_{\tau}^{i}$ is a $\hat{\rho_{i}}$-rooted topological minor of $G^{*\phi(i)}$ were $\hat{\rho}_{i}=\rho_{i}\cup\{(x_{\rm new}^{i},x_{\rm new}^{*i})\}.$ To simplify notation,
we drop indices and we denote $R^{i}_{\tau}=(R'^{i}_{\tau},\tilde{T}^{i}\cup\{x_{\rm new}^{i}\})$ and
$G^{*\phi(i)}=(G^{\prime \phi(i)},T^{\phi(i)}\cup \{x^{*\phi(i)}_{{\rm new}}\})$  
by  
$R=(R',\tilde{T}\cup\{x_{\rm new}\})$ and 
$G^{*}=(G^{\prime},T\cup \{x^{*}_{{\rm new}}\})$  
respectively. We also use $\hat{\rho}$ instead of $\hat{\rho}_{i}$ and $x_{\rm new}$ instead of $x_{\rm new}^{i}.$

We now apply the {\sl irrelevant vertex technique,} introduced in~\cite{RobertsonS-XIII} as follows. Using the algorithm from~\cite{Amir01effi} (or, alternatively, the one from~\cite{Reed92find})
one can either compute a tree-decomposition of $G^{\prime}$ of width at most $q=4\cdot (24\cdot (2\cdot g(|E(R')|)+8)(\sqrt{|T|+2})+49)$ or prove 
that no tree decomposition exists with width less than  $q/4.$
In the case where $\tw(G')\leq q$, we recall that $|E(R')|$ 
is a function of $k$ and $|T|\leq 2k.$ Consequently, there exists a function $f_{3}$ such that $\tw(G')\leq f_{3}(k)$ and the result 
follows from Lemma~\ref{lemma:msol} in the Appendix.
 
 Suppose now that $\tw(G')\geq q/4=24\cdot (2\cdot g(|E(R')|)+8)(\sqrt{|T|+2})+49.$ Applying Lemma~\ref{apost} for $r=3$ we have that ${\bf pth}_{T'}(G^{*})\geq g(|E(R')|)$ where $T'=N^{3}_{G'}(T\cup\{x_{\rm new}^{*}\}).$

We now prove that  there is a vertex $x\in V(G^{*})$
such that if $R$ is a $\hat{\rho}$-rooted topological minor of $G^{*}$ then 
$R$ is a $\hat{\rho}$-rooted topological minor of $G^{*}\setminus \{x\}.$
Let $\psi_{0}$ and $\psi_{1}$ be the functions certifying that 
$R$ is a $\hat{\rho}$-rooted topological minor of $G^{*}.$ We apply 
on $\psi_{0}$ and $\psi_{1}$ the modifications of 
the  Lemma~\ref{lem:helio} 
so that they satisfy Properties 1--4. An important consequence is that
the images of all vertices of $R$ under $\psi_{0}$ are close to
terminals in $T.$ 
Indeed, from Properties 1 and 2, all neighbors of $x_{\rm new}$ are mapped  via $\psi_{0}$ to vertices that belong in $N^{1}_{G'}(x_{\rm new}^{*}).$ Moreover, from Properties 3 and 4 it follows 
that $\psi_{0}(V(\tilde{\bf J}))\subseteq V({\bf F})\subseteq N^{2}_{G'}(x_{\rm new}^{*}).$ So far we have proved that all vertices of $R$, 
except those inside $\tilde{\bf J}$, are mapped via $\psi_{0}$ to vertices in $N^{2}_{G'}(x_{\rm new}^{*}).$
As all vertices of $R$ that are inside $\tilde{\bf J}$ belong to
$\tilde{T}$, it follows that $\psi_{0}(V(R))\subseteq T'.$

The set $L= \{\psi_{1}(e) \mid e\in E(R)\}$ is a set of paths in
$G^*.$ Let $L_1 \subseteq L$ be the set of those paths that have
length at most $2$ 
and  define $L_2 = L \setminus L_1.$  
For each path $Q \in L_2$ define its {\em interior}, denoted 
$\mbox{\rm int}(Q),$ as
the subpath of $Q$ consisting of all vertices of $I(Q).$ 
Clearly, ${\cal L}=\{  \mbox{\rm int}(Q) \mid Q \in L_2 \}$ is a
linkage in $G^{*}$ and $\pi({\cal L}) \subseteq T'.$ Consider the collection $\mathcal{C}$ 
of $T'$-concentric cycles $C_{1},\ldots,C_{g(|E(R')|)}$
certifying  the fact that ${\bf pth}_{T'}(G^{*})\geq
g(|E(R')|).$ Define the graph ${\rm \Gamma} =
\left( {\bf clos}(\Delta_{\rm in}(C_1)) \cap G^*    \right) \cup \left( \bigcup_{Q \in L_2} \mbox{\rm
int}(Q) \right).$  We have that (i) ${\cal L}$ is a linkage of ${\rm \Gamma},$  and
(ii)
$|E(R')| \geq |\pi({\cal L})|.$ 
Let $x$ be a vertex of $C_{g(|\pi({\cal L})|)}.$ 
By Proposition~\ref{seymlink} there is another linkage 
${\cal L}'$ with the same pattern as ${\cal L}$
such that all paths in ${\cal L'}$ avoid $x.$

The vertices of all  paths in $L_1$ belong to $T'$ since they are at
distance at most $3$ from $T\cup{x^*}_{\rm new}.$ 
The endpoints of all paths in $L_2$ also belong $T'$ since they are at
distance at most $2$ from $T\cup{x^*}_{\rm new}.$  Therefore 
all paths in $L_1$ and the endpoints of all paths in $L_2$ 
avoid ${\rm \Gamma}$ altogether.  We now show that all paths in $L$ can be
rerouted so that they avoid $x,$ while they remain internally
vertex-disjoint. 
The paths in $L_1$ stay the same. For the paths in $L_2,$ we only have
to reroute their interiors within ${\rm \Gamma}.$ 
This is achieved by connecting the pairs in $\pi({\cal L})$ via the
linkage 
${\cal L}'.$  After this substitution  all  paths in the updated set $L$ 
avoid $x.$  By updating $\psi_{1}$ to reflect the new interiors
of the paths in $L_2,$ we obtain that $R$ is a $\hat{\rho}$-rooted
topological minor of 
$G^{*}\setminus \{x\}.$  

Notice that $x$ can be found in linear time applying BFS starting from $T'.$ After deleting $x,$ 
we create an equivalent instance of the problem with smaller size. By recursively applying the same reduction to the new instance at most $|V(G')|$ times, at some point 
the treewidth will drop below $q$ and then we solve the problem by
applying Lemma~\ref{lemma:msol} in the Appendix 
as above.\qed
}

From Lemmata~\ref{moster} and~\ref{irelmos} we obtain  Theorem~\ref{theo:main}.

\section{Further extensions and open problems}

We chose to tackle the disjoint-paths completion problem with the  
topological restriction of having non-crossing patch edges.
A natural extension of this problem is to allow a fixed number $\xi> 0$
of crossings in the patch.  We believe that, using the same techniques, one may
devise an $f(k)\cdot n^{2}$ algorithm  for this problem as well. The only substantial difference 
is a generalization of our combinatorial result (Theorem~\ref{theo:patchsize}) under the presence of crossings. 

An interesting topic  for future work is to define and solve the disjoint-paths completion 
problem for graphs embedded in surfaces of higher genus. A necessary step in this direction
is  to extend Theorem~\ref{theo:patchsize} for the case where the face to be patched 
contains handles. 

Another issue is to extend the whole approach for the case where 
the faces to be patched are more than one. This aim can be achieved without
significant deviation from our methodology, in case the number of these faces is bounded.
However, when this restriction does not apply, the problem seems challenging and, in our opinion,
it is not even clear whether it belongs to ${\sf FPT}.$

\paragraph{Acknowledgement.}  We wish to thank the anonymous reviewers
of an earlier version of this paper for valuable comments and suggestions.

\vspace{-2mm}

{
\bibliographystyle{siam}

}

\newpage

\section*{Appendix }
\label{sec:irrelevant}

\section*{A.1 MSOL and rooted topological minors}


In this section we define a parameterized  version of Rooted
Topological Minor Testing that uses treewidth as a parameter  and we show, using Monadic Second Order Logic
(\mso), that it is in {\sf FPT.} The
corresponding algorithm will be used as a subroutine in the proof of Theorem~\ref{theo:main}. \\

{ 
\noindent\rahmen
{
	$p$-{\sc Bounded Treewidth Rooted Topological Minor Testing}
	\linie
	\textbf{Input:} a rooted graph $(H,S_{H})$, a rooted graph
$(H,S_{G})$, and a bijection $\rho \colon  S_{H}\rightarrow S_{G}.$ \\
\textbf{Parameter:} $k={\bf tw}(G)+|V(H)|$\\
	\textbf{Question:}  is  $(H,S_{H})$ is a {\em $\rho$-rooted topological minor} 
of the rooted graph $(G,S_{G})$?}
}

\begin{lemma} \label{lemma:msol}
$p$-{\sc Bounded Treewidth Rooted Topological Minor Testing} can be solved in $f(k)\cdot |V(G)|$ steps.
\end{lemma}

\proof{
Let $S_H=\{a_1,\ldots, a_s\}$, let 
$V(H)\setminus S_H=\{u_1,\ldots,u_t\}$, and
let $E(H)=\{e_1,\ldots, e_{\epsilon}\}.$\\
Let $\tau_s=\{\text{VERT},\text{EDGE}, I, c_1,\ldots c_s\}$ 
be the vocabulary of graphs (as incidence structures) with $s$ constants. 
We give an $\mso[\tau_s]$ formula $\phi_{H,S_H}$ such that for every graph
$G$ with $S_G=\{b_1,\ldots, b_n\}$ and $\rho(a_i)=b_i$ we have 
that
	$(G,b_1,\ldots,b_s)\models \phi_{H,S_H} \iff 
	(H,S_{H})\text{ is a }\rho$-rooted topological minor of $(G,S_{G}).$
	
Let \textit{path}$(x,y,Z)$ be the $\mso[\tau_s]$ formula stating that
$Z$ is a path from $x$ to $y.$ (This can be easily done by saying that $Z$ is a set of edges
with the property that for every vertex $v$ incident to an edge in $Z$, the vertex $v$ is either incident to
precisely two edges in $Z$, or $v$ is incident to one edge in $Z$ and $v=x$ or $v=y.$
Finally, we can express that the path $Z$ is connected.) \\
Let 
\begin{align*}
	\phi_{H,S_H}:=&\exists Z_1 \ldots Z_{\epsilon}\exists x_1 \ldots x_t\\ 
	& \big(
	\bigwedge_{i\neq j; i,j\leq t}x_i\neq x_j\wedge
	\bigwedge_{e_i=\{u_k,u_{\ell}\}\subseteq V(H)\setminus S_H}\textit{path}(x_k,x_{\ell},Z_i)\;\wedge\\
	&\bigwedge_{\substack{e_i=\{u_k,a_{\ell}\}\\ x_k\in V(H)\setminus S_H\\
	a_{\ell}\in S_H }}\textit{path}(x_k,c_{\ell},Z_i)
	\wedge\bigwedge_{e_i=\{a_k,a_{\ell}\}\subseteq S_H}\textit{path}(c_k,c_{\ell},Z_i)\;\wedge\\
	&\bigwedge_{i\neq j;\;i,j\leq \epsilon}\exists y\big(
	(\text{VERT}y \wedge 
	\text{`}y\text{ is incident to an edge in }Z_i\\
	& 
	\text{\hspace{5cm} and to an edge in }Z_j\text{'})\\
	&\to (\bigvee_{i=1}^s y=c_i \vee\bigvee_{i=1}^ty=x_i ) \big)\big).
\end{align*}
The constants $c_i$ are interpreted by the $b_i$, hence they make sure that $a_i$ is mapped to $b_i$, and 
Condition 1 of rooted topological minors is satisfied.
In addition we make sure that every edge of $H$ is mapped to a path in $G.$
Finally we make sure that Condition 3 is satisfied.
(The statement `$x$ is incident to an edge in $Z_i$ and to an edge in $Z_j$' 
can be easily expressed in $\mso.$)

Observe that the length of the formula $\phi_{H,S_H}$ only depends on
$H.$ Hence by Courcelle's Theorem \cite{Courcelle90them,CourcelleM92mon} there is a computable function $f_{1}$ such that 
$p$-{\sc Bounded Treewidth Rooted Topological Minor Testing} can be solved in time $f_{1}({\bf tw}(G)+|V(H)|)\cdot |V(G)|.$\qed

\end{document}